\documentclass[12pt,prl,aps,nofootinbib]{revtex4}
\usepackage{graphicx,epsf,epsfig}
\newcommand{\beq}{\begin{equation}}
\newcommand{\eeq}{\end{equation}}

\newcommand{\be}{\begin{eqnarray}}
\newcommand{\ee}{\end{eqnarray}}
\def\eq#1{{Eq.~(\ref{#1})}}
\def\fig#1{{Fig.~\ref{#1}}}
\newcommand{\as}{\alpha_S}

\begin{document}
\title{CGC, QCD Saturation and RHIC data\\
( Kharzeev-Levin-McLerran-Nardi  point of view )
}

\author{Eugene Levin}

\affiliation{HEP Department, School of Physics,\\
Raymond and Beverly Sackler Faculty of Exact Science,\\
Tel Aviv University, Tel Aviv 69978, Israel\\
}

\begin{abstract}
\centerline{Talk at Workshop:``Focus on Multiplicitioes", Bari, Italy,
17-19 June,2004.}

~
~

We are going to discuss ion-ion and deuteron - nucleus  RHIC
data
and show that they support,  if not more, the idea of the new QCD phase:
colour glass condensate with saturated parton density.
\end{abstract}.

\maketitle

\section{Introduction}
\begin{figure}[hbp]
\begin{minipage}{10cm}
\begin{center}
\includegraphics[width=0.60\textwidth]{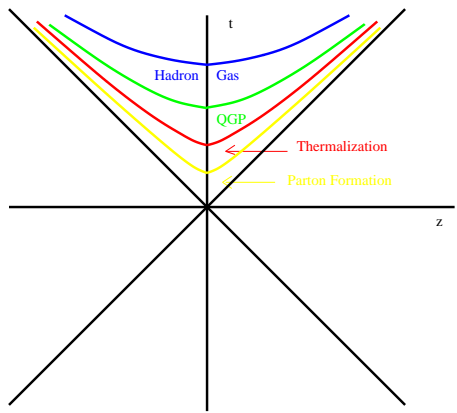}
\end{center}
\end{minipage}
\begin{minipage}{5cm}{
\caption{The space-time picture of the ion-ion collision with the main stages of the evolution in the final state: parton 
formation , thermalization, Quark-Gluon Plasma (QGP) and hadron gas.}}
\end{minipage}
\label{pic}
\end{figure}

 We used to think that nucleus-nucleus interaction is so complicated that 
the distance between the experimental data and underlying microscopic 
theory: QCD, is so large that we are not able to give any interpretation 
of the data based on QCD. My goal is to show that this wide spread opinion 
is just wrong. I hope to convince you that the nucleus-nucleus collisions 
provide  such an information on the initial stage of the process, parton 
formation (see 
\fig{pic}),  which could be (and even has been)  very essential in the 
disscussion of the new QCD phase: colour glass condensate (CGC).

\begin{figure}[htbp]
\begin{center}
\includegraphics[width=0.70\textwidth]{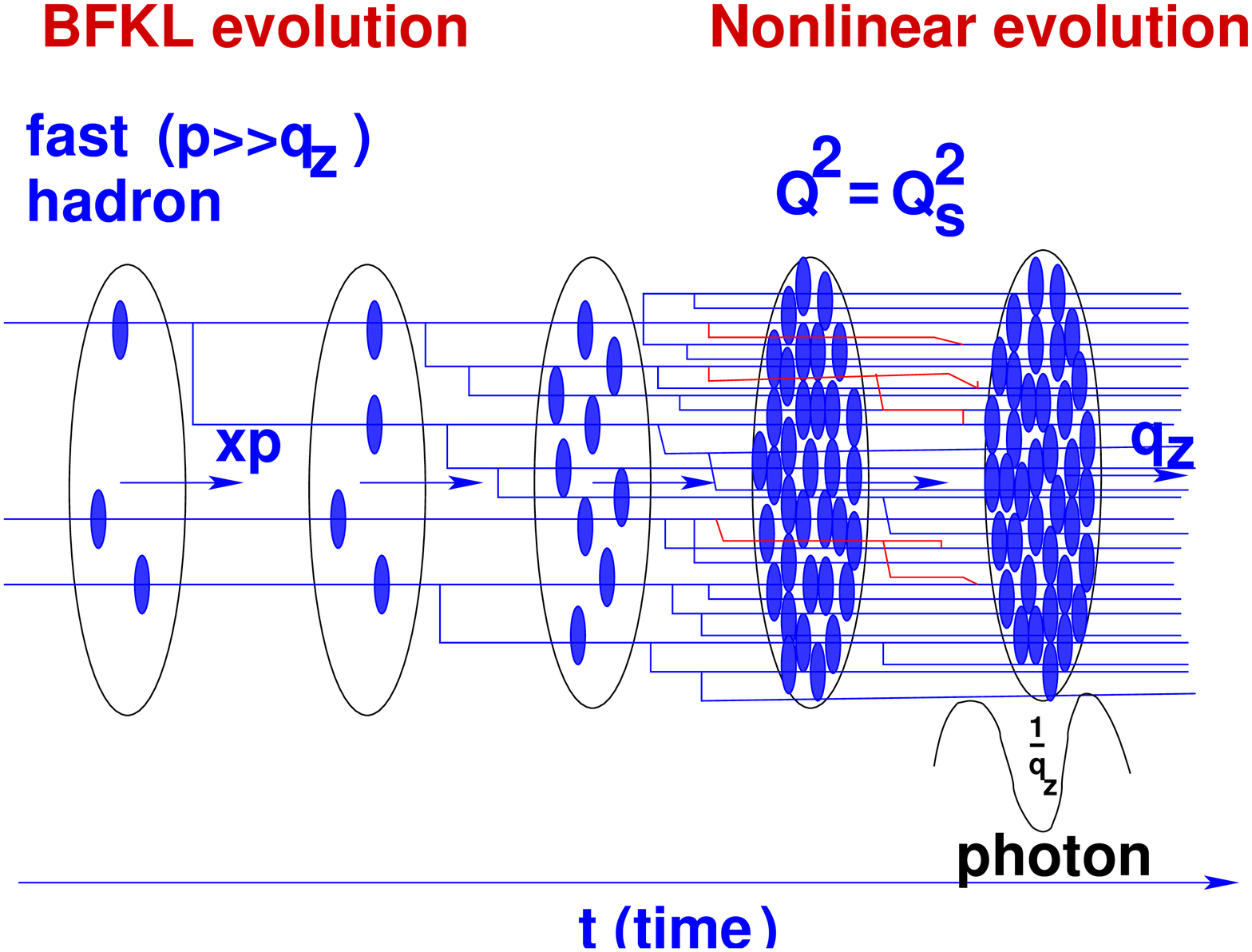}
\end{center}
\caption{Space-time picture for the deep inelastic scattering.}
\label{picdis}
\end{figure}

\begin{figure}[htbp]
\begin{center}
\includegraphics[width=0.65\textwidth]{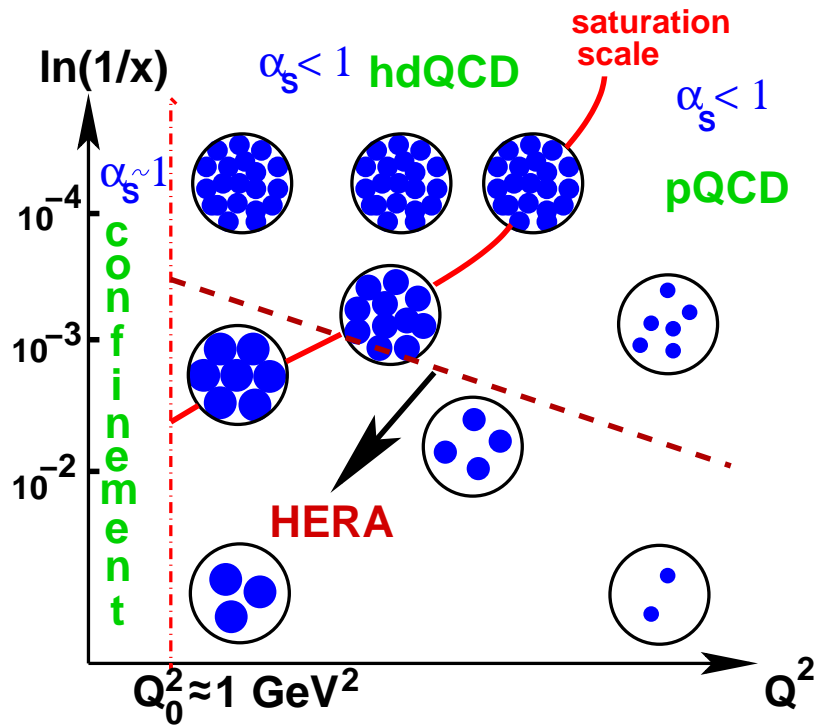}  
\end{center}
\caption{The distribution of partons in the transverse plane.} 
\label{sat-pic}
\end{figure}

The main prediction of the CGC is the fact that the parton density in CGC region is saturated reaching a maximal 
value\cite{GLR,MUQI,MV}.
The space-time picture of the QCD saturation is shown in \fig{picdis}.  Let us consider a virtual photon, with virtuality 
$Q$, in the rest frame 
(Bjorken frame) where the photon is the standing wave which interacts with the parton (colour dipoles) with size 
$r\,\approx\,1/Q$.  In the beginning of 
our process we have a small number of partons of this size but at late time the number of partons steeply increases. At some 
moment of time the partons start to populate densely in the proton and filled the whole proton disk (see \fig{sat-pic}).

The estimate t of the value of a new scale: saturation scale which relates to the size of the parton when the partons 
started to populate densely (the critical curve in \fig{sat-pic}) we introduce the  packing factor:
\beq \label{PF}
 \mbox{P.F.} \,\,\, \equiv\,\, \,\kappa\,\,\,\, =
\sigma_{parton}
\,\times\,\rho \,\,\, \approx
\frac{3\,\pi^2 \alpha_S}{2
Q^2_s(x)}\,\times\,
\frac{xG(x,Q^2_s(x))}{\pi\, R^2}
\eeq
The saturation scale is the solution to the equation:
\beq \label{QS}
\kappa (Q_s(x)) \,\, =\,\,1
\eeq

\section{Theory status}

\begin{figure}[hbtp]
\begin{center}
\includegraphics[width=0.80\textwidth]{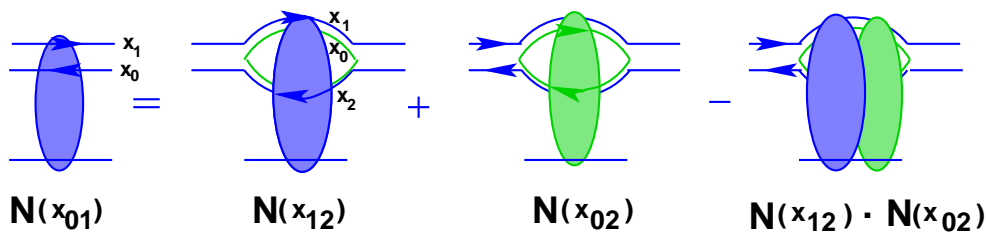}
\end{center}
\caption{The Balitsky-Kovchegov equation.}
\label{bk}
\end{figure}

The scope of this talk does not allow us to discuss theory of the high parton density systen but, I firmly believe, that a 
reader should know,  our approach is based on the first principles of the microscopic theory (QCD).  Despites  rather 
complicated technique,  that we have to use approaching this regime, the theoretical ideas, which we based upon, are very 
transparent and can be easily explained and digested.  For the diluted system of partons the main process is the emission of 
gluon and this process leads to the famous DGLAP evolution equation. However, when the density of partons increases the 
processes of recombination, which are proportional to the square of density ($\rho^2$), should enter to the game. The 
competition between 
emission ($\propto \rho$), which increases   the number of partons, and recombination ($\propto \rho^2$), which diminishes 
this 
number, results in the equilibrium density. The phenomenon of approaching the maximal density we call `parton 
density saturation' and the phase of QCD with saturated density is the colour glass condensate. The evolution equation which 
describes the saturation phenomenon is a non-linear equation \cite{GLR,MUQI,MV} which final form was  found by Balitsky and 
Kovchegov \cite{BK}.  This equation is so beautiful that I decided to present it here despites the lack of room.

\beq \label{BK}
\frac{\partial N(y,\vec{x}_{01},\vec{b})}{\partial y}\,\, = \,\frac{C_F
\as}{2 \pi^2}\,\int\,d^2 x_{2}\, \frac{x^2_{01}}{x^2_{02}\,\,x^2_{12}}
\,   \left(  2 N(y,\vec{x}_{12},\vec{b} -
 \frac{1}{2} \vec{x}_{02})\,   \right.
\eeq
$$
 \left.   - \,
N(y,\vec{x}_{01},\vec{b})\,\, -\,\,N(y,\vec{x}_{12},\vec{b} - \frac{1}{2}
\vec{x}_{02})\,N(y,\vec{x}_{02},\vec{b} - \frac{1}{2} \vec{x}_{12})
  \right)
$$
where $x_{ij}$ are the sizes of dipoles (see \fig{bk}).

\eq{BK} shows that the degrees of freedom at high energy are the colour dipoles rather than quark and gluon themselves 
\cite{MUCD}. The physical meaning of $N$ is the dipole-target amplitude. The equation says that the change of this amplitude
from $y$ to $y + d y$ where $ y=\ln t = \ln(1/x)$ and $x$ is the Bjorken variable fir DIS. is equal to the probability for 
the incoming dipole to decay into two dipoles. These two dipoles can interact with the target separately and simultaneously. 
The  simultaneous interaction should be taken with the negative sign which reflects the shadowing effect or  accounting for 
the the recombination processes.

The Balitsky-Kovchegov equation at the moment is the main theoretical tool for in many applications of the CGC dynamics 
despites it's very approximate nature. It plays a role of the mean field approach in this problem. 

\begin{figure}[hbtp]
\begin{center}
\includegraphics[width=0.80\textwidth]{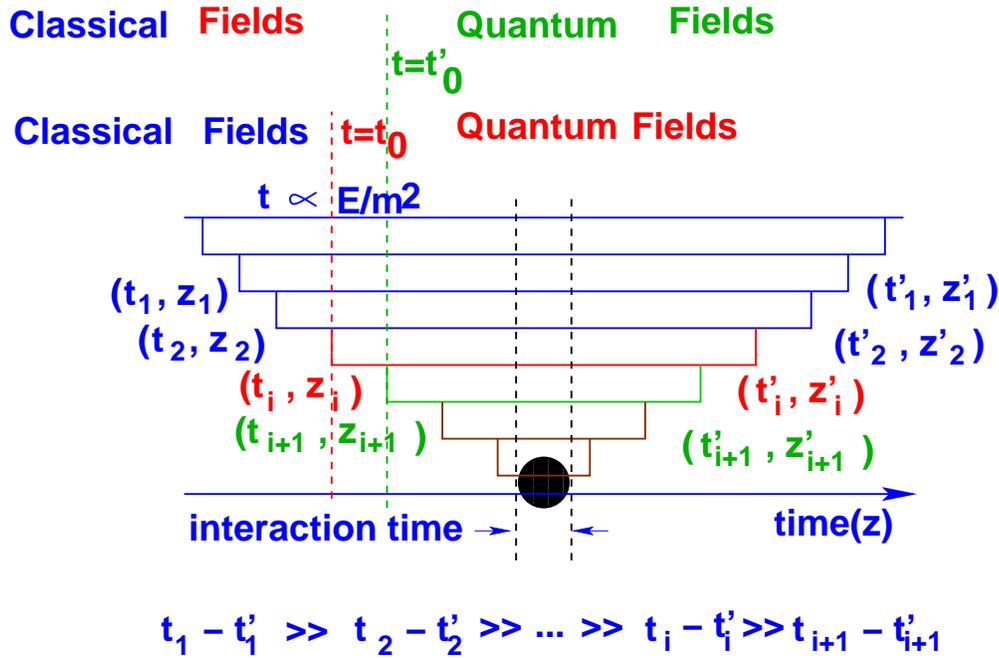}
\end{center}
\caption{The  space-time structure of high energy interaction in QCD and the renormalisation Wilson group approach (JIMWLK 
equation) to high parton density QCD.}
\label{cgctime}
\end{figure}

It is very important to understand that we have a more general approach than the mean field one. This approach is based on 
the space-time structure of the high energy interaction in QCD (see \fig{cgctime}). The idea of this approach is very 
transparent. Let us start with emission of the parton at time shown by the first dotted line in \fig{cgctime}. All parton 
with high energies were created  a long before this moment. The main idea \cite{MV}  is that these partons can be treated 
classically  while the parton emitted at this moment can be  described by QCD.  Moving the moment of time (see the second 
dotted line in \fig{cgctime} we should include the gluon produced in $(t_i,z_i)$ into the parton system which we consider 
classically, but a new gluon emitted in $(t_{i+1},z_{i+1})$ shall be treated in full QCD. Since the description should be the 
same in these two moments ($ t_i $ and $t_{i+1}$ we have a constraint which leads to the equation. The realization of this 
program is rather complicated as well as technique that is required to understand the resulting equation but the physical 
basis for the JIMWLK equation \cite{JIMWLK} is very simple.   

The last remark that I would like to make in this brief theoretical introduction is related to the theoretical input which 
we will use in our description of the experimental data. It turns out that we need to know the solution to the non linear 
equation only in vicinity of the  saturation scale together with the general understanding of the behaviour of the dipole 
amplitude in the saturation domain. Since the unitarity itself leads to $N \,\rightarrow \,1$ the different equations can 
only specify the character of approaching this maximal value. The description of the RHIC data do not depend on the details 
of this approaching. The behaviour near to the saturation boundary can be found just knowing the solution to the linear 
evolution equation. 
\section{Comparison with the data: general approach.}
I believe that the most important thing is to stipulate clearly what kind of assumptions we are making  applying the general 
theory to an interpretation of the experimental data at accessible energies and typical distances reached by the experiment. 
Indeed, the theory was formulated  for the dense parton system while the experimental data exist rather in the transition 
domain where density is not very high but not low as well.  In our KLMN approach\cite{KN,KL,KLN,KLM,KLND,KLMAZ} we used three 
main assumptions which we will 
have to reformulate at the end of the talk.  
\begin{enumerate}
\item \quad  At $x \,=\,x_0\,\approx\,\,10^{-1}$ we have the McLerran-Venugopalan
model for the inclusive production of gluons with the saturation scale
$Q^2_s(x_0) \,\,\approx\,\,1\,GeV^2$ \cite{MV};
\item \quad
The RHIC region of $x \,\approx\,10^{-3}$ is considered as the low $x$ region in which
$\as\,\ln(1/x) \,\approx\,1$ while $\as \,\ll\,1$.  This is not a principal
assumption, but it makes the calculations much simpler and more transparent;

\item \quad
We assume that the interaction in the final state does not change significantly
the distributions of particles resulting from the very beginning of the process.
For hadron multiplicities, this may be a consequence of local parton hadron duality,
or of the entropy conservation. Therefore multiplicity measurements are extremely
important for uncovering the reaction dynamics.  However, we would like to state clearly that wee do not claim
that interaction in the final state are not important. We rather consider the CGC as the initial condition for the
interaction in the final state. 

\end{enumerate}

\section{Comparison with the data: HERA data.}
We would like to make three statements about deep inelastic scattering data from HERA: (i) models that incorporate the 
parton density saturation (see for example Refs. \cite{GW,BGK,BGLM,KOTE}) are able to describe the HERA data; (ii) the 
solution to the Balitsky-Kovchegov 
equation leads to the good description of the experimental data on the DIS  structure functions \cite{GLMF2,IIM}; and (iii)  
the 
gluon structure function extracted from the fit of the experimental data is so large that the packing factors reaches the 
value of unity.

\begin{figure}[h]
\begin{minipage}{18pc}
\includegraphics[width=18pc]{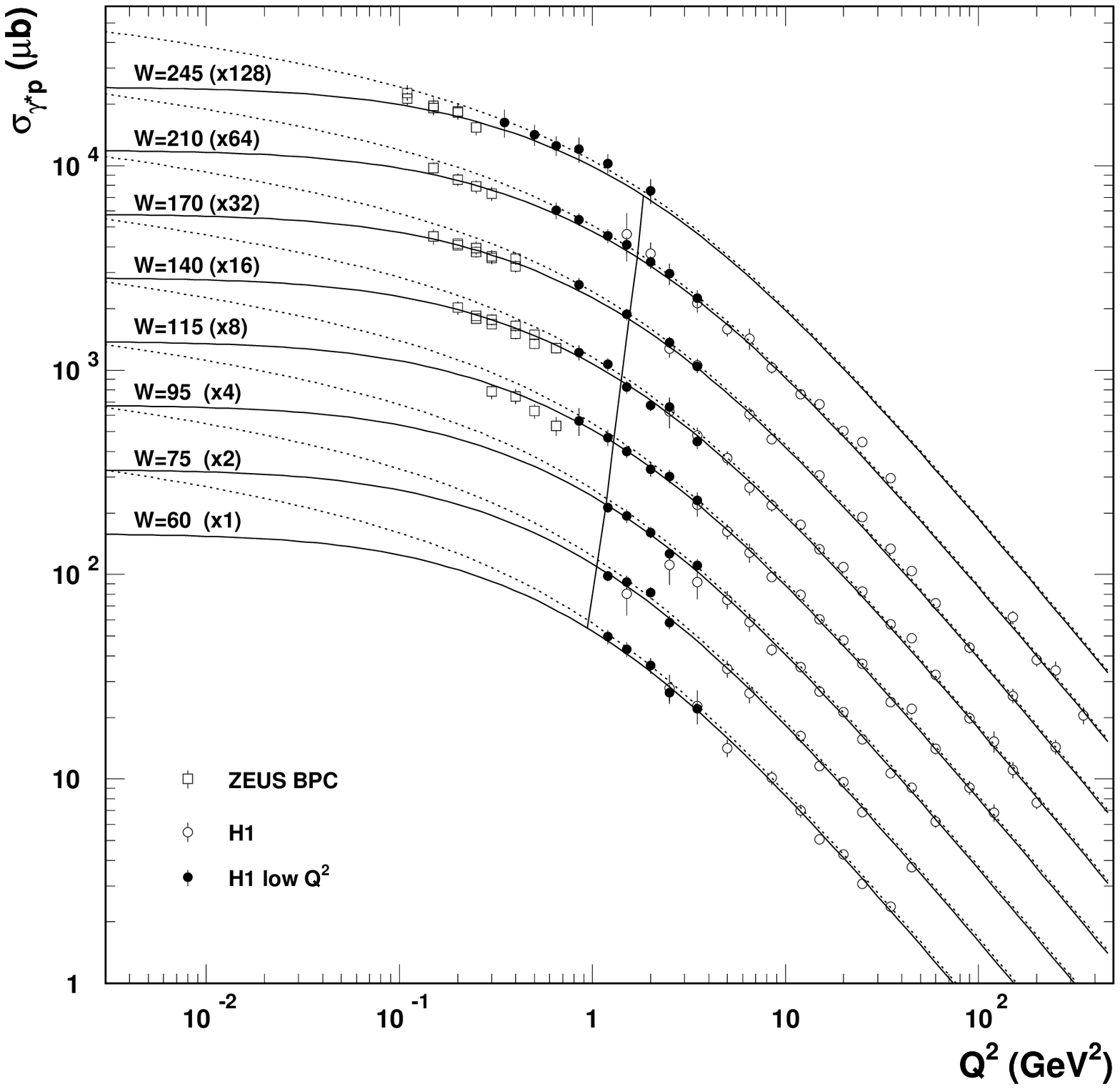}
\caption{\label{gw1} $\gamma^* p$ total cross sections in the 
Golec-Biernat 
and Wuesthoff model}

\end{minipage}\hspace{2pc}%
\begin{minipage}{18pc}
\includegraphics[width=18pc]{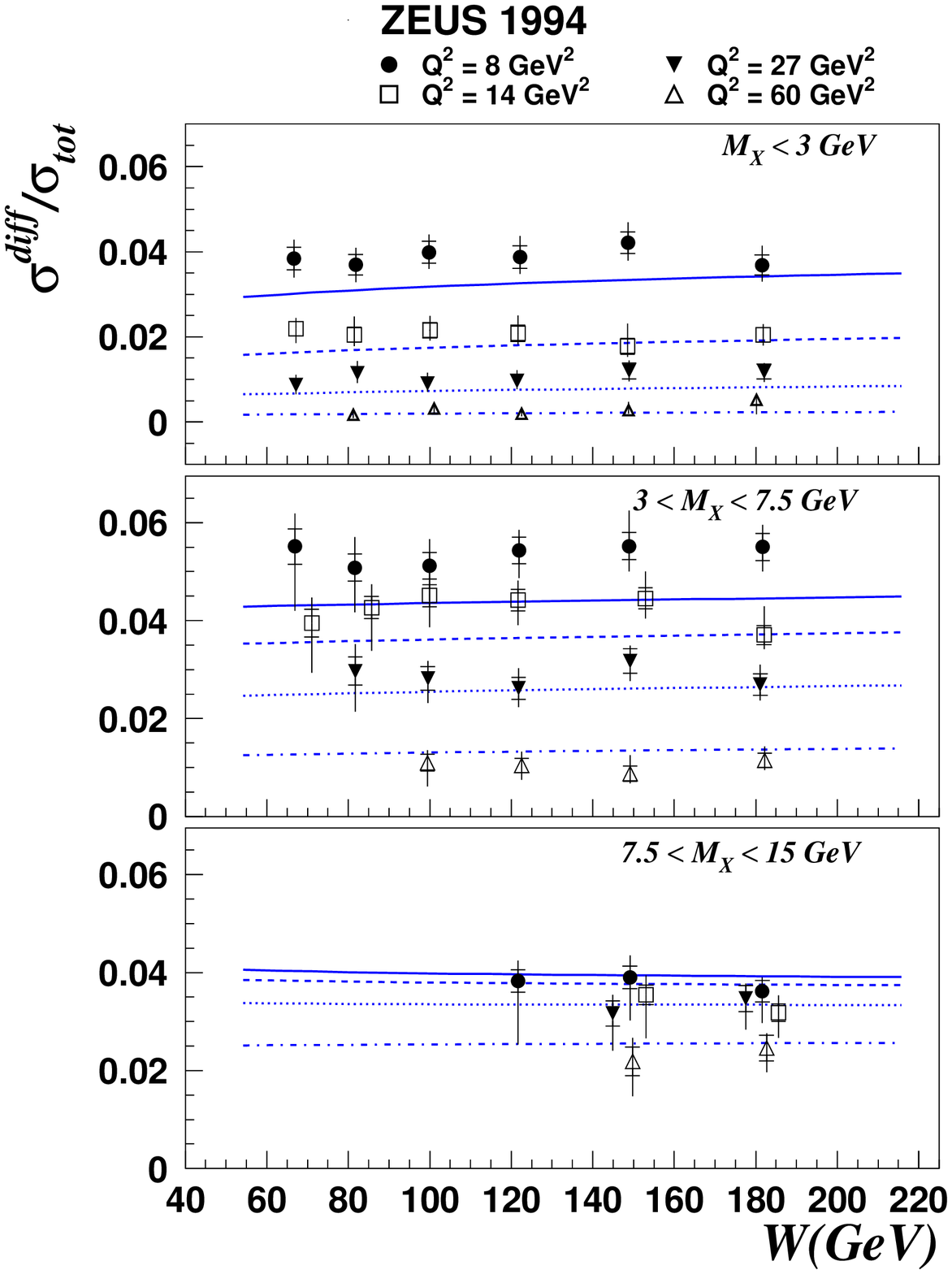}
\caption{\label{gw2} DIS diffraction  in the Golec-Biernat and Wuesthoff 
mode .}
\end{minipage}
\end{figure}

The set of figures illustrate  this our point of view.  \fig{gw1} and \fig{gw2} show the comparison with the experimental 
data the simple Golec-Biernat and Wuesthoff model in which the dipole-proton cross section is written in the form
\beq \label{XSDP}
\sigma(x,r_{\perp})\,\,\,=   
\,\,\,\sigma_0\,\,\left(\,\,1\,\,\,-\,\,\,
\exp(\,-\,\frac{r^2_{\perp}}{R^2(x)}\,)\,\,\right)
\eeq
with
  $ R^2(x)\,\,\,=\,\,\,1/Q^2_s(x)$ , where 
\beq \label{QSGW}
Q^2_s(x)\,\,=\,\,Q^2_0 \,\left(\,\frac{x}{x_0}\,\right)^{-\lambda}
\eeq
The parameters were chosen from the fit of the experimental data and they are:
$\sigma_0$\,\,   =\,\,    23.03\,mb\,;\,\,
$\lambda$\,\,  = \,\,  0.288 \,;\,\,
$x_0$\,\,   = \,\,  $ 3.04\,10^{-4}$;\,\,
$ Q^2_0$\,\,  =\,\, $ 1\,GeV^2$.

\begin{figure}[hbtp]
\begin{center}
\includegraphics[width=0.70\textwidth]{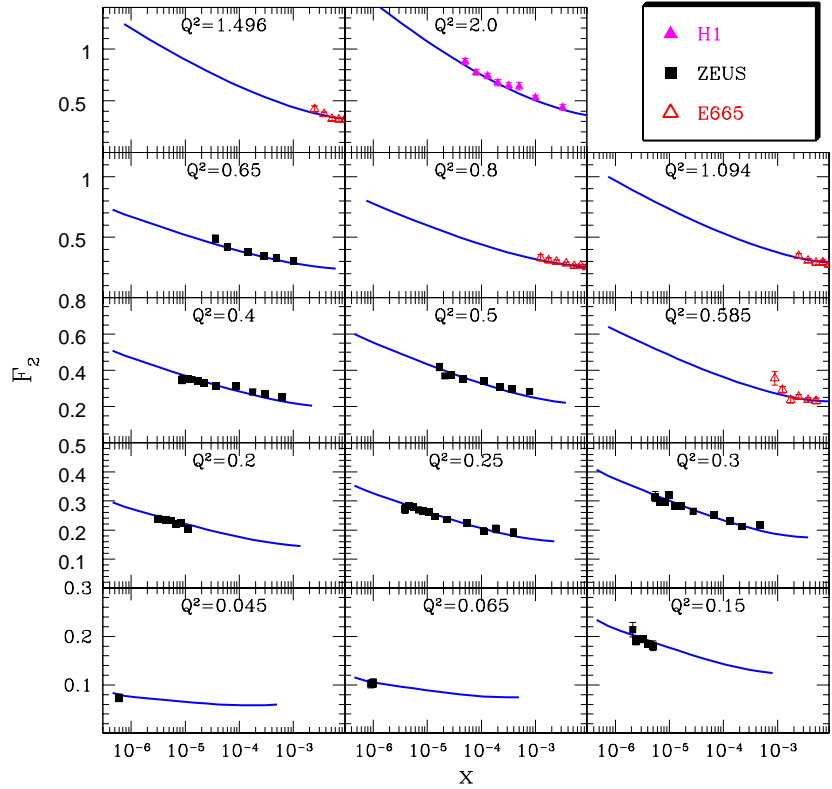}
\end{center}
\caption{$F_2$ as result of solution to the  Balitsky-Kovchegov equation. Picture is taken fron Ref. \cite{GLMF2}}
\label{f2bk}
\end{figure}

\fig{f2bk} gives the example how the Balitsky-Kovchegov equation is able to describe the deep inelastic structure function.

\begin{figure}[hbtp]
\begin{minipage}{10cm}
\begin{center}
\includegraphics[width=0.50\textwidth]{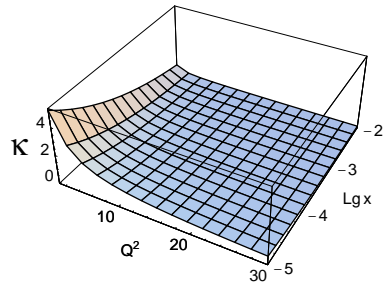}          
\end{center}     
\end{minipage}
\begin{minipage}{5cm}
\caption{Packing factor $\kappa$ (see \eq{PF}) from the HERA data. }
\label{kappa}
\end{minipage}
\end{figure}

\fig{kappa} gives the packing factor calculated in one of the parametrization for the extracted gluon structure function.

\section{Multiplicity at RHIC}
\begin{figure}[hbtp]
\begin{center}
\includegraphics[width=0.80\textwidth]{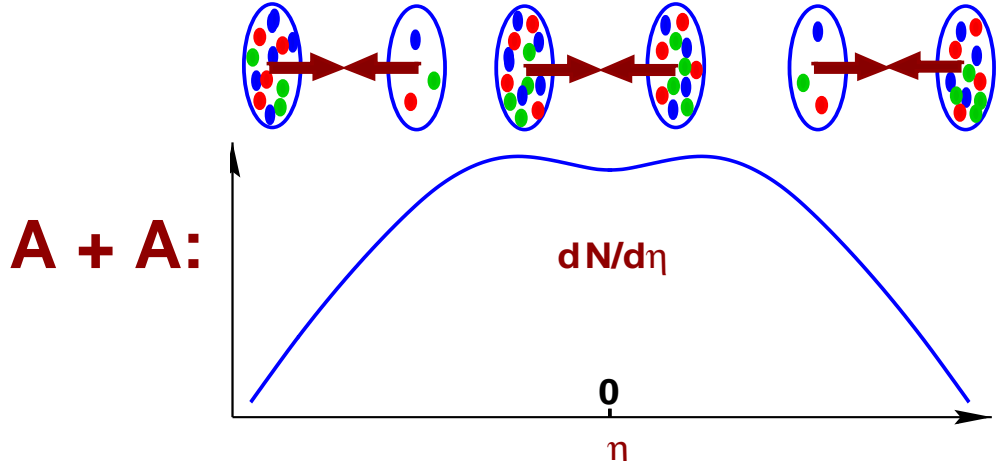}
\end{center}     
\caption{Ion-ion collision in the CGC approach. }
\label{aapic}
\end{figure}

Multiplicity is the most {\it  reliable} test of our
approach since
the { \it  third assumption}  is {\it  correct} due to entropy
conservation. The physical picture of ion-ion collision in the CGC approach  is given by \fig{aapic}.
In central region of rapidities we have a collision of two dense system of partons while in the fragmentation regions
one system of partons is rather diluted while the second one turns out to be more dense than at $\eta=0$.

To characterize the ion-ion collision we use several observables, namely, centrality cut, which related too typical value of 
the impact parameters; number of participants ($N_{part}$) , which shows the number of proton taking part in interaction; and 
the number of 
collisions ($N_{coll}$) . All these observables are discussed in details in Nardi's talk at this conference and I will assume 
that you are 
familiar with them.  Here, I only want to formulate the simple rule: if a physical observable  $
\propto\,\,\,N_{part}$ this observable can be described by \,\,\,{\it soft
physics} while observables $\propto\,\,\,N_{coll}$  are certainly related to typical hard processes, which can be treated in 
perturbative QCD.  
Soft physics at scale {\it larger} than $1\,\,GeV$
is {\it  a  great surprise}. In simple words the saturation in CGC is a claim that soft physics starts at $p_t 
\,\leq\,Q_s(x)$ and $Q_s$ could be large at high energies.

The key relation in the CGC is that the saturation scale is proportional to the number of participants or
\beq \label{QSA}
Q^2_s(A)\,\,=\,\,\frac{3 \pi^2}{2}\,\alpha_S(Q^2_s)
xG_N(x,Q^2_s)\, \frac{\rho_{part}}{2}
\eeq
and 
\beq \label{QSAPA}
S_A \cdot Q^2_s(A)\,\,  \propto\,\,N_{part}
 \eeq

In the case of the ion-ion collisions we have actually two saturation scales in colliding ions. In the simple Golec-Biernat 
and Wuesthoff model they have the following dependence on energy $W$ and  rapidity $y$:
\be 
Q^2_{min}\,\, =\,\, \,Q^2_s(A;W=W_0;y=0)\,\left(\frac{W}{W_0}
\right)^\lambda\,e^{-\,\lambda\,y}\,\,;\label{QSMIN}\\
Q^2_{max}\,\, = \,\,
\,Q^2_s(A;W=W_0;y=0)\,\left(\frac{W}{W_0}
\right)^\lambda\,e^{\lambda\,y}\,\,;\label{QSMAX}
\ee

 In  our  papers 
we use
 the following formula for the inclusive production \cite{GLR,GM}:
\beq \label{INCR}
E {d \sigma \over d^3 p} = {4 \pi N_c \over N_c^2 - 1}\ {1 \over p_t^2}\  \times
\, \int^{p_t} \, d k_t^2 \,\varphi_{A_1} (x_1, k_t^2)\,\,\varphi_{A_2} (x_2,(\vec{p}_t -  \vec{k}_t)^2)
\eeq
where $x_{1,2} = (p_t/\sqrt{s}) \exp(\mp y)$ and
$\varphi_{A_1,A_2} (x, k_t^2)$ is the unintegrated gluon distribution of a nucleus ( for the case of the proton
one
of $ \varphi_{A}$ should be replace by $ \varphi_{p}$.)
This distribution is related to the gluon density by   
\be \label{XGPHI}
xG(x,Q^2) \,\,=\,\,\int^{Q^2}\,\,d \,k^2_t \, \varphi(x, k_t^2).
\ee
The multiplicity can be calculated using the following formula:
\be \label{MUG}
{dN \over dy} & = & {1 \over S}\ \ \int d p_t^2 \left( E {d \sigma \over d^3 p} \right)\,\,\,=\,\,\,
               {4 \pi N_c \as \over N_c^2 - 1}\,\frac{1}{S}  \nonumber \\
        &\times& \,\,\int {d p_t^2 \over p_t^2}\ \left(
\,\varphi_{A_1}(x_1, p_t^2)\,\, \int^{p_t} \, d k_t^2 \
 \varphi_{A_2}(x_2, k_t^2) \,\,+\, \,\varphi_{A_2}(x_2, p_t^2)\,\, \int^{p_t} \, d
k_t^2  \ \varphi_{A_1}(x_1, k_t^2) \, \right)  \nonumber \\
&=& {4 \pi N_c \as \over N_c^2 - 1}\,\,\frac{1}{S}\,
\,\int^{\infty}_0\,\,\frac{
d\,p^2_t}{p^4_t}\,\,x_2G_{A_2}(x_2,p^2_t)\,\, x_1G_{A_1}(x_1,p^2_t)\,\,,
\ee
   where we integrated by parts and used \eq{XGPHI}.
In the KLMN treatment we assumed a simplified form of $xG$, namely,
\be \label{XGSAT}
xG(x;p^2_t)\,\,=\,\,\left\{\begin{array}{l}\,\,\,\, {\kappa \over \as(Q^2_s)}\,
S\,
p^2_t\,\,(1\, -\, x)^4\,\,\,\,p_t\,<\,Q_s(x)\,\,; \\ \\
\,\,\,\, {\kappa \over \as(Q^2_s)}\,
S\,Q^2_s(x)\,\,(1\, -\, x)^4\,\,\,\,p_t\,>\,Q_s(x)\,\,;
\end{array}
\right.
\ee
Inserting \eq{XGSAT} into \eq{MUG} we obtain a simple formulas:
\beq \label{MUF}
 \frac{d N}{d y} \,\,=\,\,Const \,\,S_A\,\,\,Q^2_{s, min}\,\ln\left(
\frac{Q^2_{s,min}}{\Lambda^2_{QCD}}\right) \times  \left(\,1\,\,+\,\,
\frac{1}{2} \ln\left(\frac{Q^2_{s,min}}{Q^2_{s, max}}\right)\,( 1 -
\frac{Q_s}{\sqrt{s}} \,e^y  )^4 \right) 
\eeq

\begin{figure}[hbtp]
\begin{center}
\includegraphics[width=0.90\textwidth]{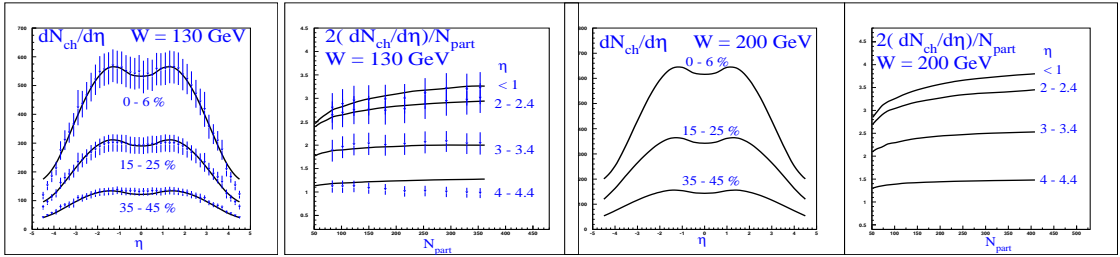}
\end{center}
\caption{Rapidity dependence of measured multiplicity. }
\label{mu1}
\end{figure}

\begin{figure}[htbp]
\begin{minipage}{18pc}
\includegraphics[width=18pc]{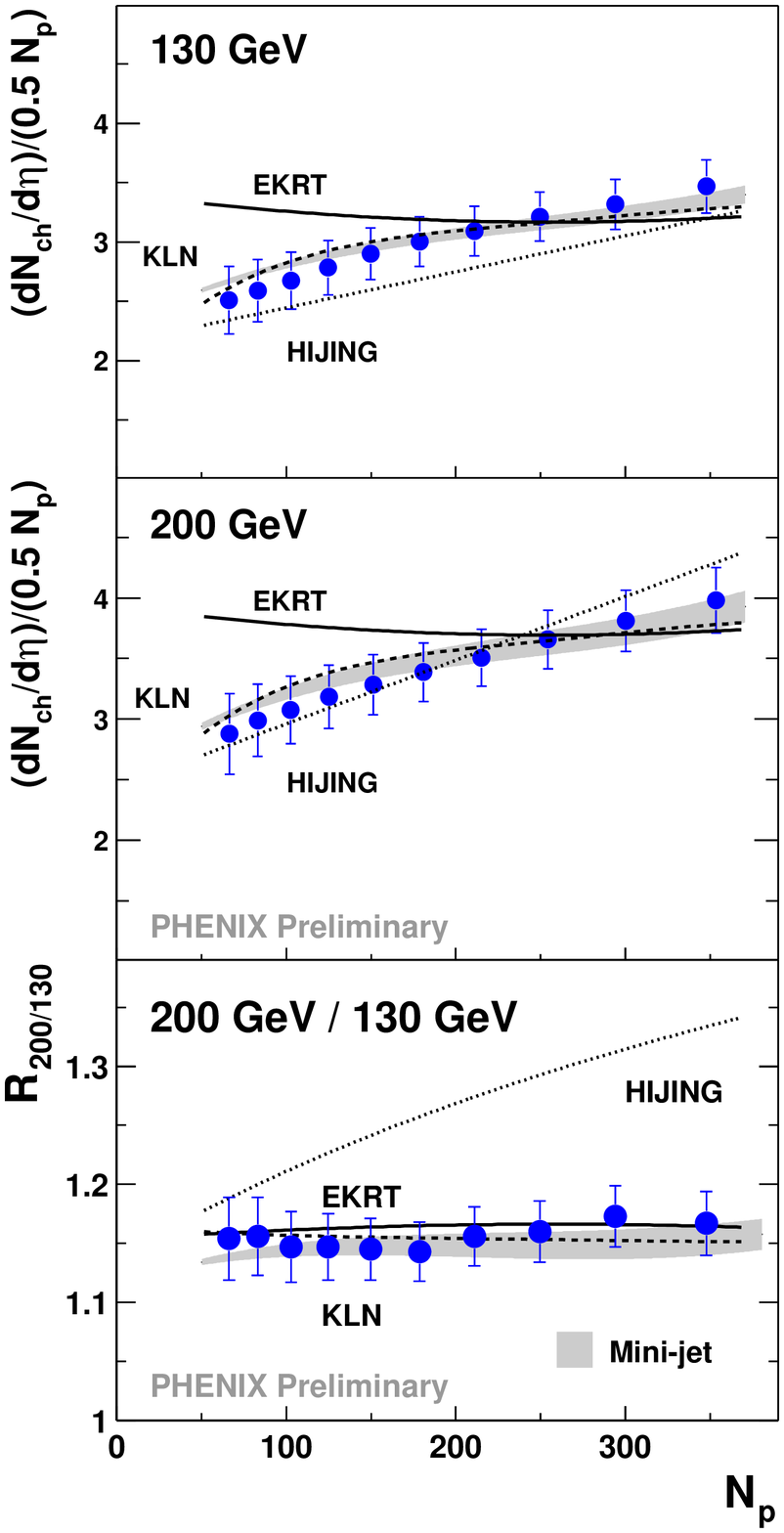}
\end{minipage}\hspace{2pc}%
\begin{minipage}{18pc}
\caption{\label{mu2} PHENIX data for $N_{part}$ dependence of the 
multiplicity distributions. Our calculations are labeled 
by `KLN' in this figure.}
~\\
~\\
\includegraphics[width=18pc]{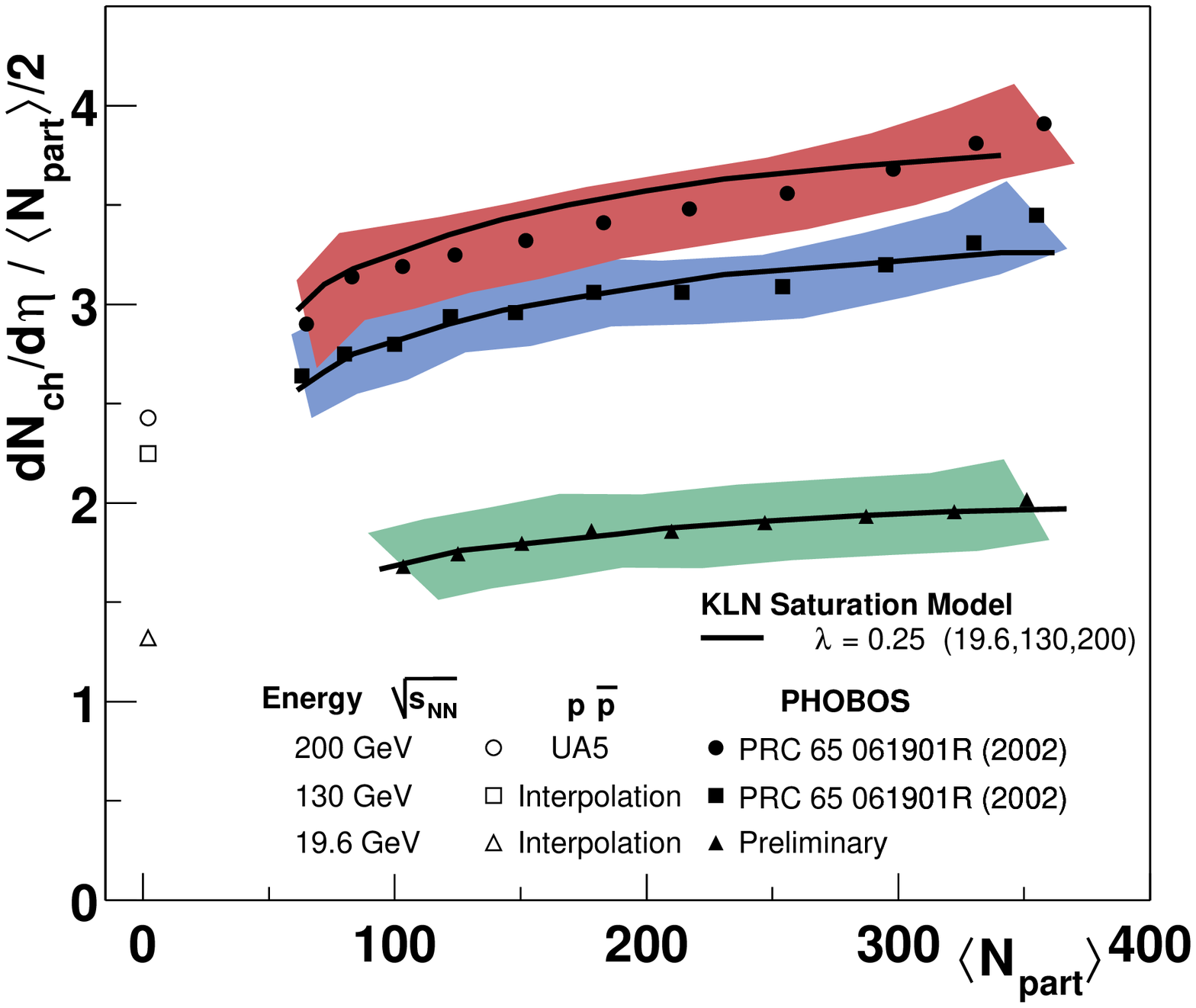}
\caption{\label{mu3} PHOBOS data on energy and $N_{part}$ dependence of 
the multiplicity distributions.
Our calculations are labeled by `KLN saturation model' in this figure. $\lambda$ determines the energy and rapidity 
dependence of the saturation scale ( see \eq{QSMIN} and \eq{QSMAX})}
\end{minipage}
\end{figure}

This simple formula with the saturation scales determined by \eq{QSMIN} and \eq{QSMAX} describes quite well the rapidity, 
energy and $N_{part}$ dependence of the multiplicity (see \fig{mu1}, \fig{mu2} and \fig{mu3}) that have been measured at 
RHIC \cite{Brahms,Phenix,Phobos,Star}.

\begin{figure}[htbp]
\begin{minipage}{18pc}
\includegraphics[width=18pc]{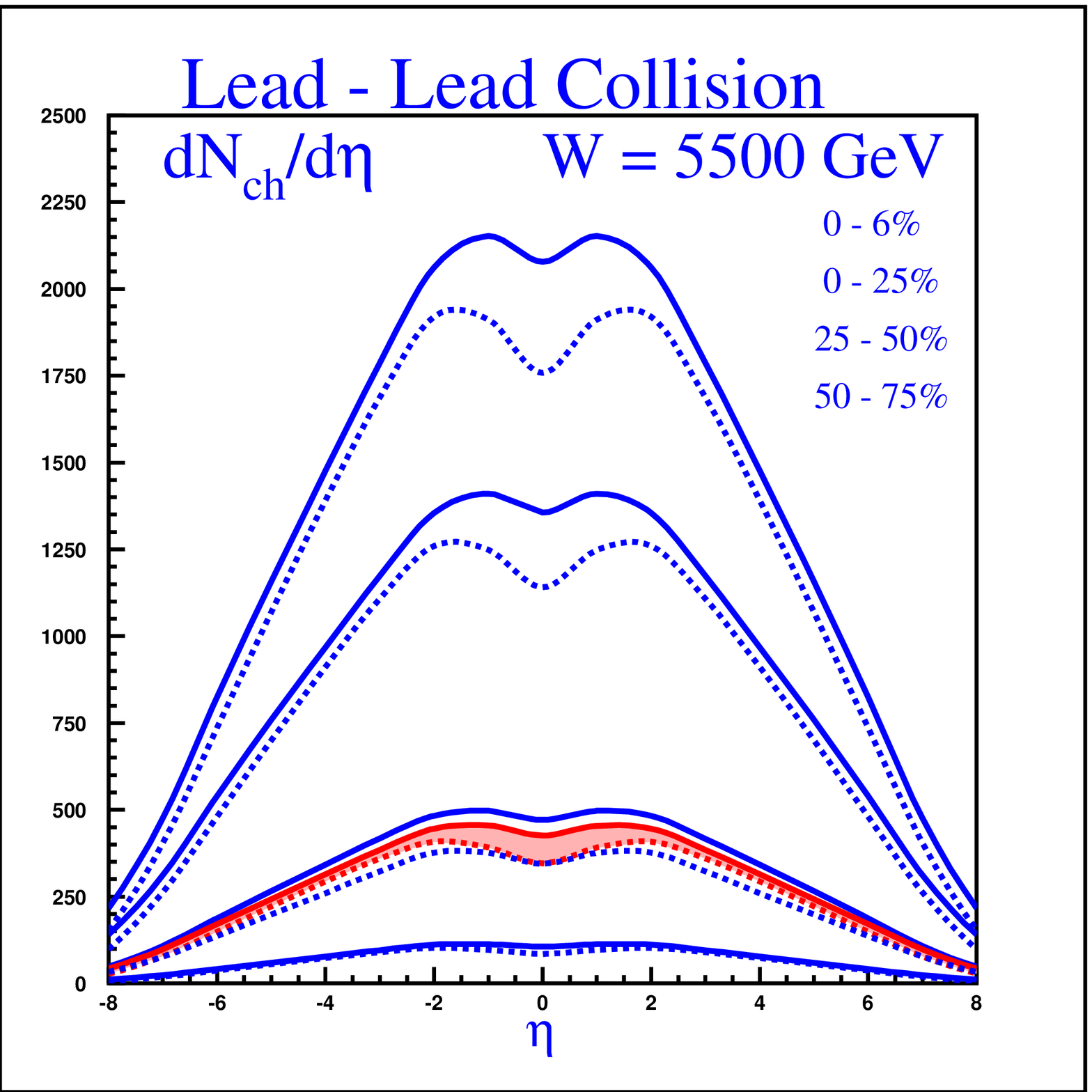}
\caption{\label{pred1} The KLN prediction for the LHC energy. The full 
curves correspond the Golec-Biernat and Wuesthoff 
model for the energy dependence of the saturation scale. The dotted line describe the prediction if energy dependence of 
the saturation scale is determined by the running QCD coupling. See Ref. \cite{KLNLHC} for more details.}
\end{minipage}\hspace{2pc}%
\begin{minipage}{18pc}
\includegraphics[width=18pc]{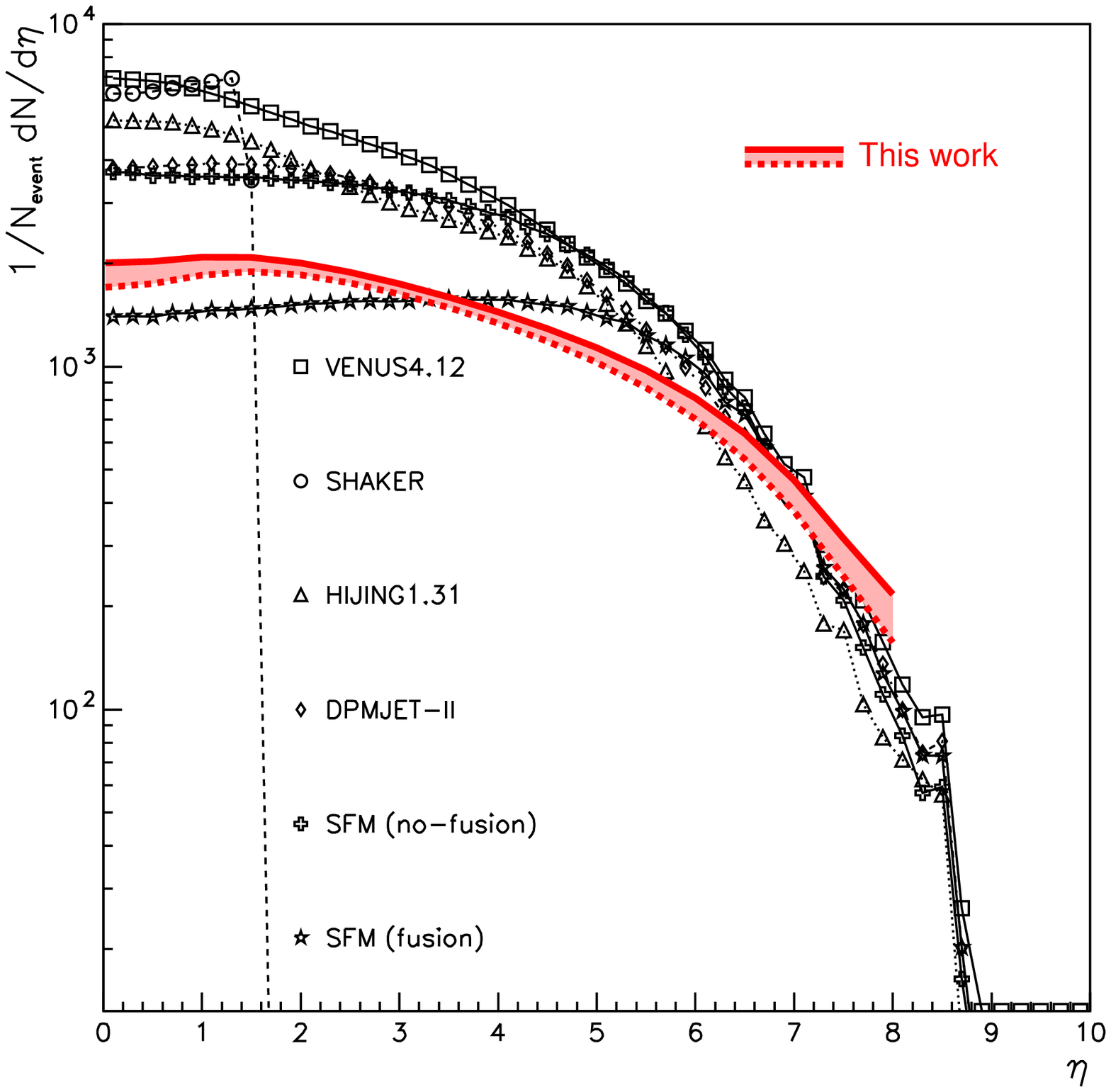}
\caption{\label{pred2} Comparison our prediction with other models available on the market (see Ref. \cite{REVAP}.}
~\\
~\\
\end{minipage}
\end{figure}

Using \eq{MUF} we are able to predict the multiplicity distribution at the LHC energy \cite{KLNLHC}. The systematic errors in 
our predictions mostly stem from the uncertainties in the energy behaviour of the saturation scale (see Ref. \cite{KLNLHC} 
for 
details).  The predictions are shown in \fig{pred1} and \fig{pred2}. One can see that we expect rather low value of the 
multiplicity in comparison with the majority of other approaches.

\begin{figure}[h]
\begin{center}
\includegraphics[width=0.50\textwidth]{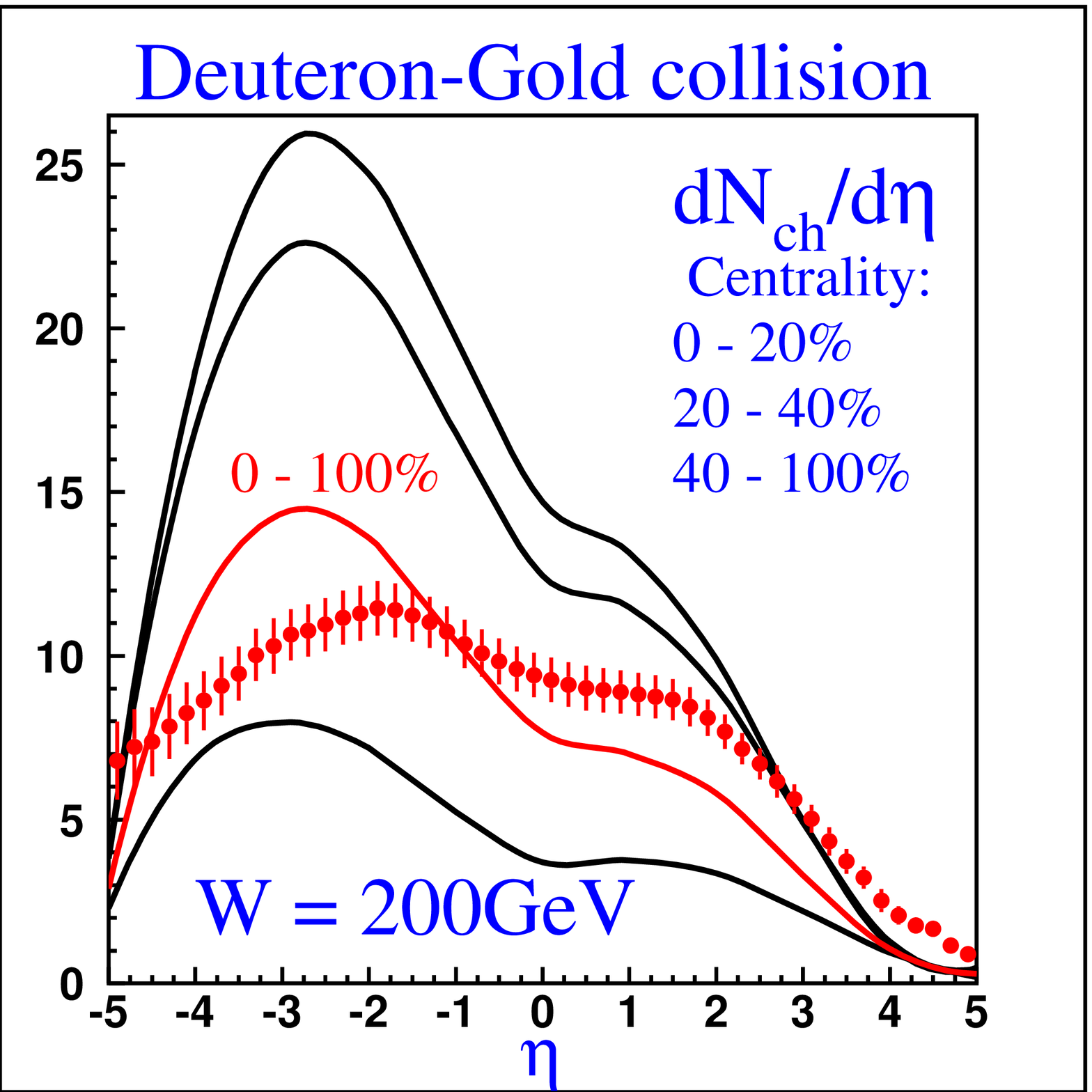}
\end{center}
\caption{Our prediction for the deuteron-gold collision at RHIC. }
\label{de1}
\end{figure}

\begin{figure}[h]
\begin{center}
\includegraphics[width=0.90\textwidth]{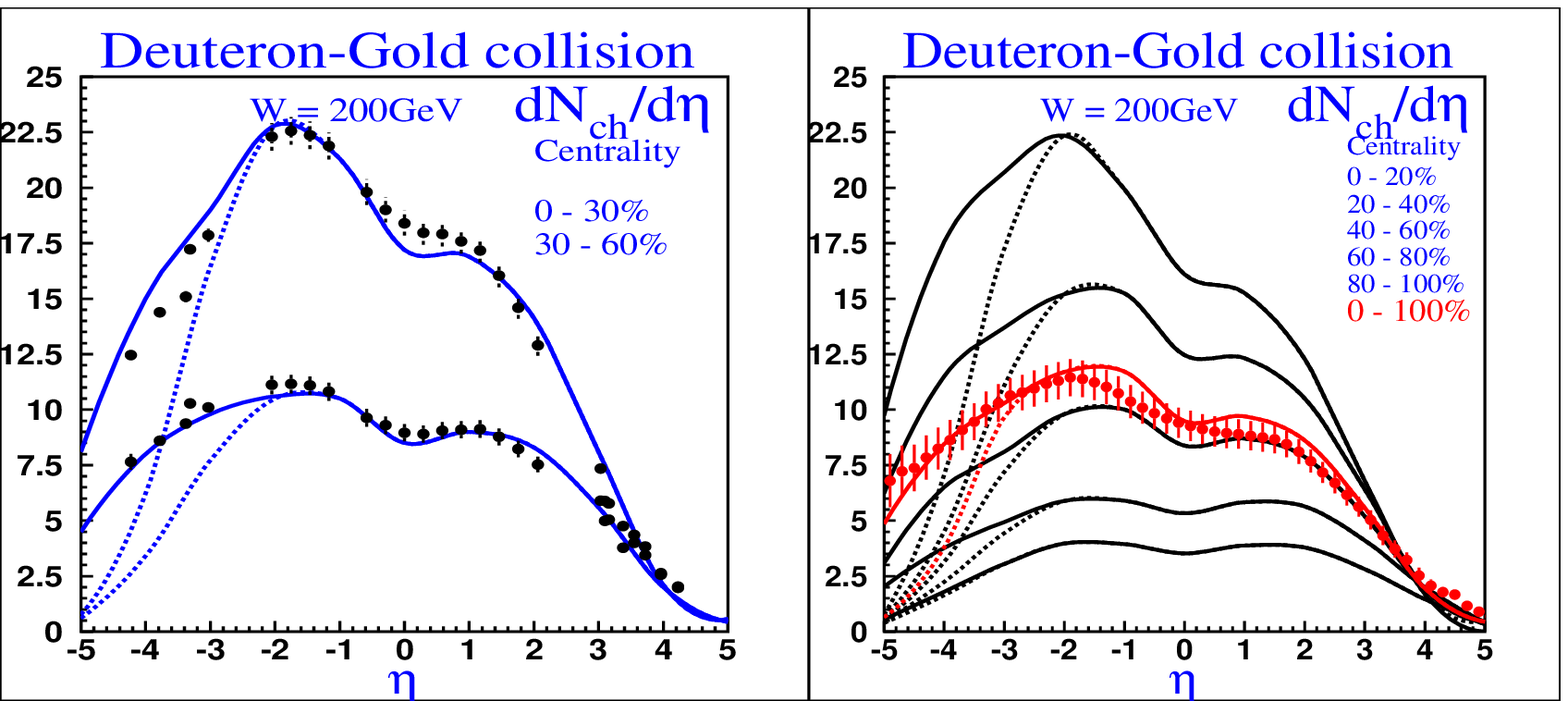}
\end{center}
\caption{Our final fit of the data  for the deuteron-gold collision at RHIC. The left figure for BRAUMS data while the right 
one for the PHONOS data  }
\label{de2}  
\end{figure}

The same formula could be generalized to describe the deuteron-nucleus collisions \cite{KLND}.  In \fig{de1} and \fig{de2} we 
show our predictions for the multiplicity distribution for deuteron-gold collisions. \fig{de1} was published before RHIC 
data and one can see that our predictions do not agree with the experimental quite well. Therefore, we had a dilemma: the 
approach is not correct or some phenomenological parameters were chosen incorrectly. Fortunately, the second is the case. 
Indeed, we found that we have to change several parameters. It turns out that the number of participants calculated by us 
(see Nardi's talk) does not agree with the number of participants that experimentalist calculated using the Glauber Monte 
Carlo. In new comparison we use the experimental value for $N_{part}$. Second, we treated incorrectly the events with the 
number of participants in the deuteron less than 1. The third change was in  the value of the saturation momentum of the 
proton. We have to take it just the same as in the Golec-Biernat and Wuesthoff model while in our old fit we took it by 30\% 
less. After these corrections the comparison is not bad (see \fig{de2}) but we cannot describe the fragmentation region of 
nucleus.


\begin{boldmath}
 \section{$N_{part}$ scaling at RHIC for ion-ion collisions}
\end{boldmath}

\begin{figure}[h]
\begin{minipage}{18pc}
\includegraphics[width=18pc]{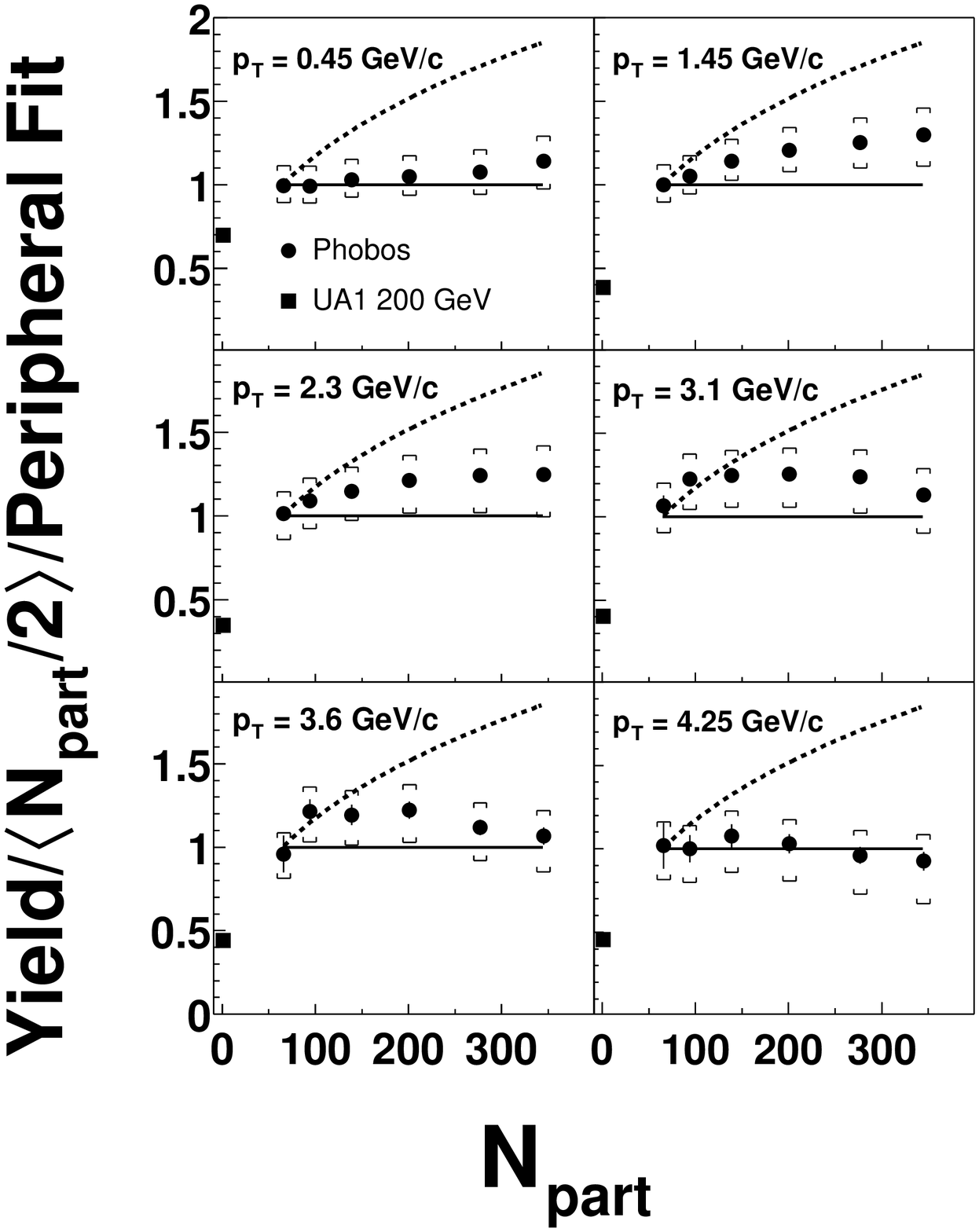}
\end{minipage}\hspace{2pc}%
\begin{minipage}{18pc}
\caption{\label{npsc1}   $N_{part}$ scaling for gold-gold collision (PHOBOS data)}
~\\
~\\
\includegraphics[width=18pc]{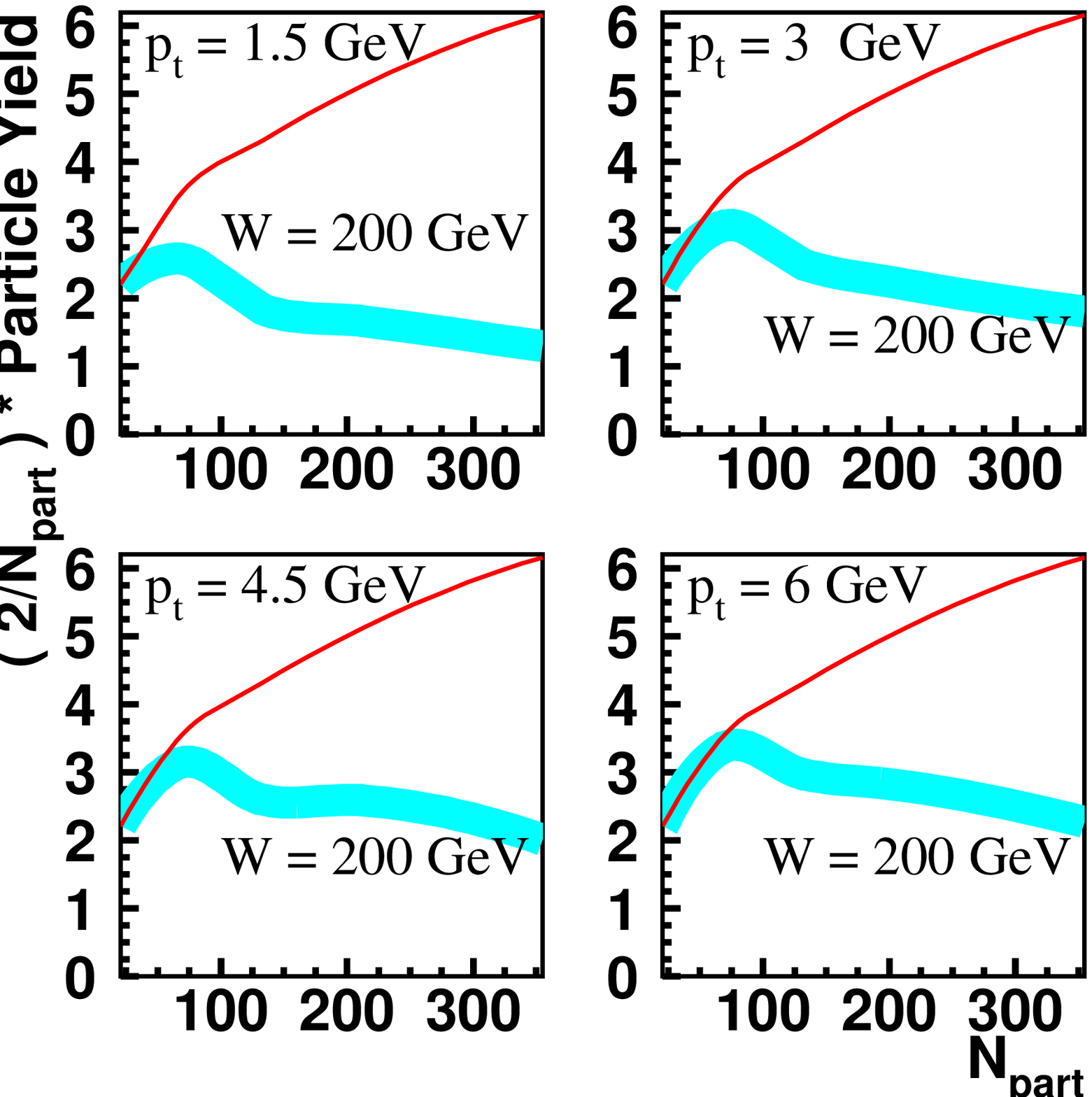}
\caption{\label{npsc2} $N_{part}$ scaling in the CGC approach ( KLMN saturation model )\cite{KLM}}
\end{minipage}
\end{figure}

One of the most interesting results from RHIC, I believe, is so called $N_{part}$ scaling. As you can see in \fig{npsc1} 
experimental data show that $dN/d y d ^2p_t\,\propto \,N_{part}$ at all values of $p_t$ up to $p_t \,\approx\,5\,GeV$. As we 
have discussed this fact is a strong indication that even at sufficiently large momenta the mechanism close to the `soft 
physics' works. This scaling behaviour has a very simple explanation in the CGC approach, which is based on three 
observations \cite{KLM}:
\begin{enumerate}
\item \quad  Geometrical scaling behaviour in the wide region outside of the saturation domain\cite{IIMGS}:\,$\phi(x, p^2_t) 
\,\,=\,\,\phi(Q^2_s(x)/p^2_t)$;
 \item \quad The anomalous dimension of gluon density is
$\,\, \approx \,\,\,\frac{1}{2}$ for $ Q^2_s(x)\,\,\,
 <\,\,\,p^2_t\,\,\,
<\,\,\,\frac{Q^4_s(x)}{\Lambda^2\,N_{part}}$;
 \item \quad
For wide range of distances the saturation scale for a
nuclear target\,\,\, $ \propto\,\,\, N_{part}\, \,\,
\approx \,\,\,A^{\frac{1}{3}}$
\end{enumerate}

\begin{figure}[h]
\begin{minipage}{10cm}
\begin{center}
\includegraphics[width=0.70\textwidth]{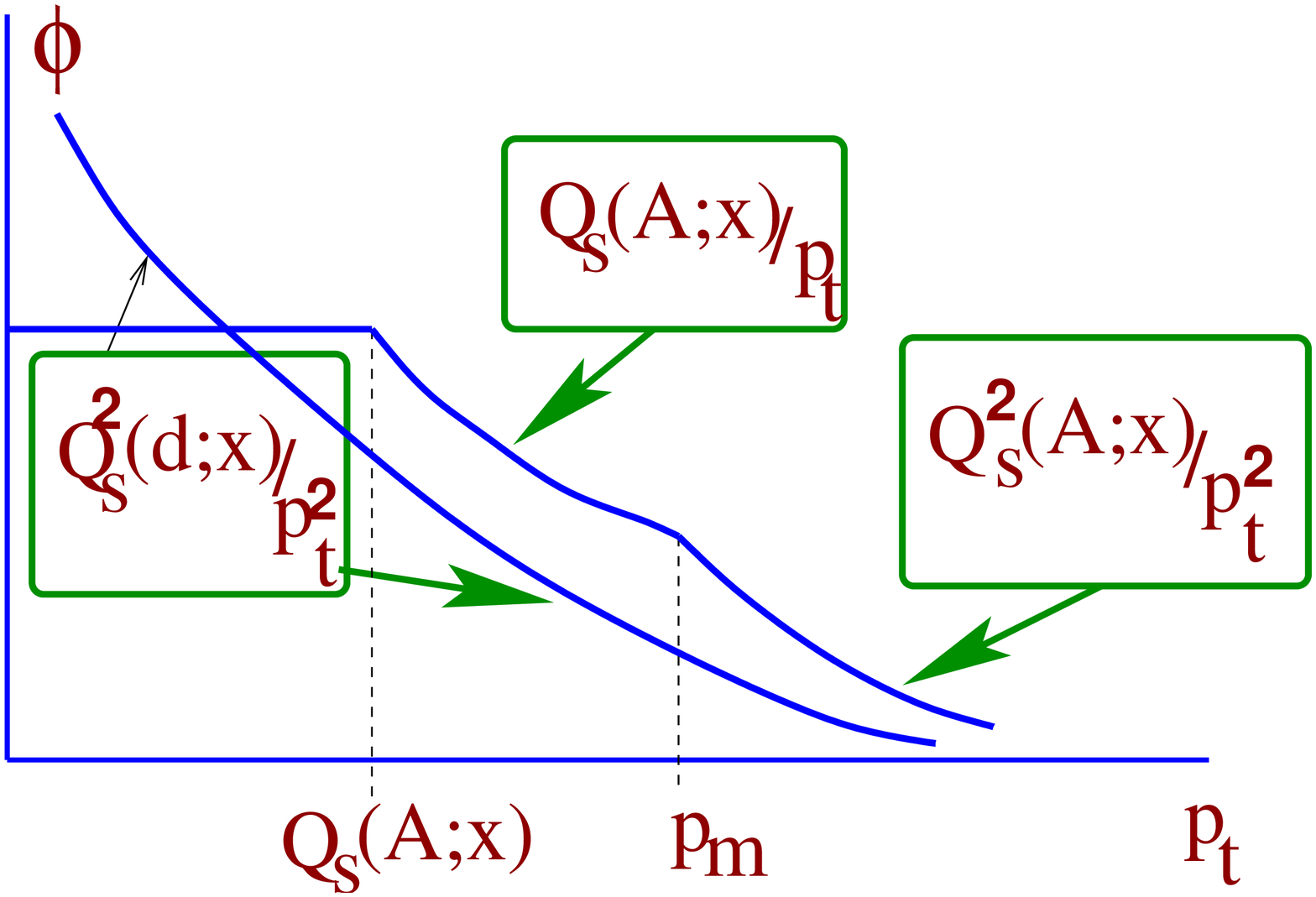}
\end{center}
\end{minipage}
\begin{minipage}{5cm}{
\caption{The behaviour of functions $\phi$ for nucleus and proton in the CGC approach taking into account the anomalous 
dimension.}}
\end{minipage}
\label{phisc}
\end{figure}

The behaviour of function $\phi$ looks as it is shown in \fig{phisc}
Using such $\phi$-s we obtain \cite{KLM} for ion-ion collisions:
\beq \label{NPSC1}
\frac{1}{S_A}E {d \sigma \over d^3 p}\,\,\,  =\,\,\,\frac{d N}{d y
d^2 p_t}
\,\,\,
\propto\,
\,\,\,S_A Q^2_s/p^2_t\,\,\, \rightarrow\,\,\,N_{part}/p^2_t
\eeq

More precise calculation shown in \fig{npsc2} lead to $N_{part}$ scaling and can describe the experimental data.

\begin{figure}[h]
\begin{minipage}{18pc}
\centering
       \mbox{{\epsfig{figure=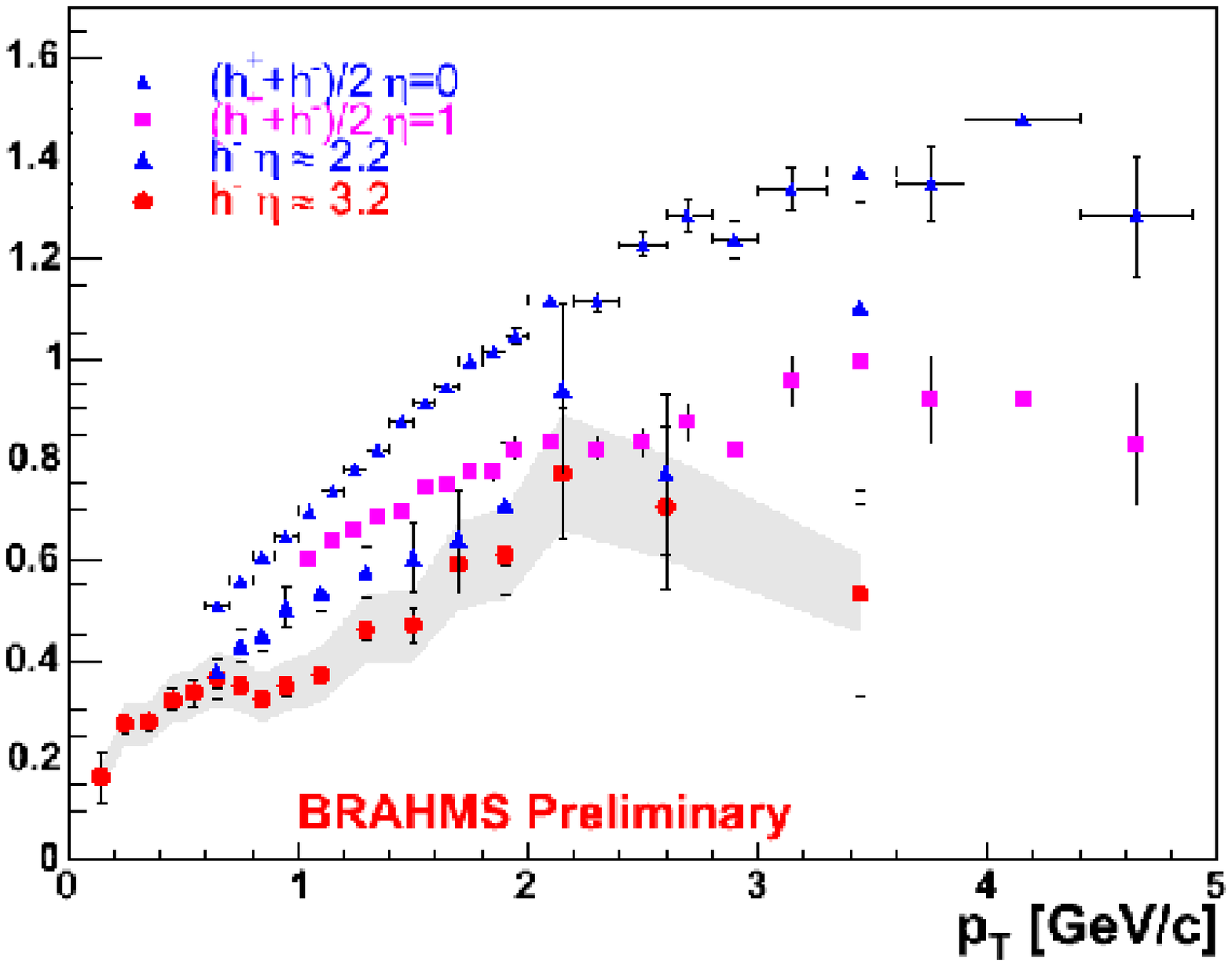,
        width=0.80\textwidth,angle=270.}}}
\caption{\label{npda1}  Deuteron-nucleus collisions in central and nucleus fragmentation region. )}
\end{minipage}\hspace{2pc}%
\begin{minipage}{18pc}
\centering
       \mbox{{\epsfig{figure=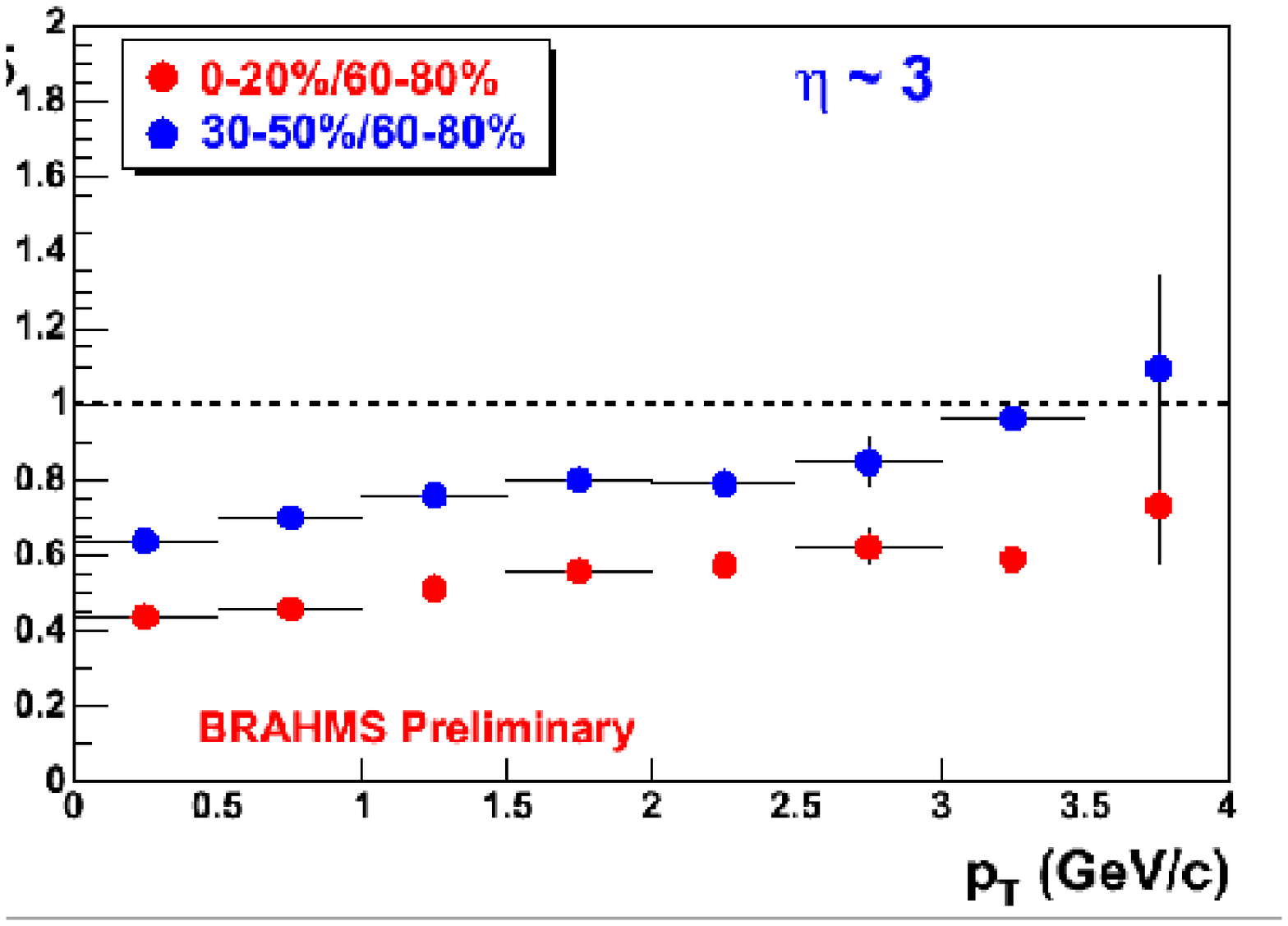,
        width=0.80\textwidth,angle=270.}}}
~\\
\caption{\label{npda2} BRAHMS data in forward region}
\end{minipage}
\end{figure}

\begin{boldmath}
 \section{$N_{part}$ scaling and suppression in deuteron-nucleus collisions}
\end{boldmath}
Using the behaviour of functions $\phi$ shown in \fig{phisc}  for deuteron-nucleus collision we obtain:
\beq \label{NPSC2}
\frac{1}{S_A}E {d \sigma \over d^3 p}\,\,\,  =\,\,\,\frac{d N}{d y
d^2 p_t}
\,\,\,
\propto
\,\,\,S_D Q_s(A)Q_S(D) /p^2_t\,\,\, \rightarrow\,\,\,\sqrt{N_{part}}/p^2_t
\eeq
\eq{NPSC2} means that we expect a suppression in comparison with wide spread opinion that $d N/du d^2 p_t$ should be 
proportional to $N_{part}$ \cite{KLM}.  The experiment shows that we do not have such a suppression in the central and 
nuclear 
fragmentation regions of 
rapidities but it certainly exists for forward region (see \fig{npda1} and \fig{npda2}). The first result needs an 
explanation the second one is a great success of the CGC approach.
 
It should be stressed that the fact that the ration at $\eta=0$ and at low $p_t$ is much smaller than 1 itself is a strong 
argument in favour of the CGC since in ordinary (Glauber) approach it should be equal to unity. As far as I know there is no 
other explanation of this suppression.

\begin{figure}[h]
\begin{minipage}{10cm}
\begin{center}
\includegraphics[width=0.90\textwidth]{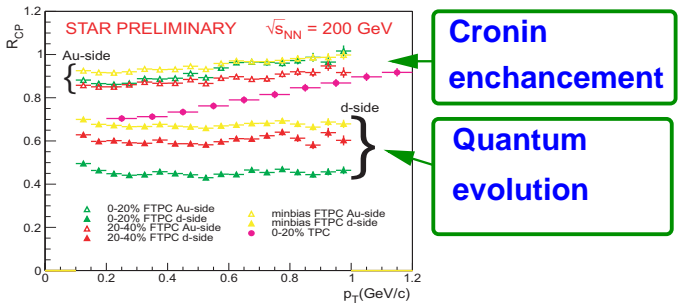}
\end{center} 
\end{minipage}
\begin{minipage}{5cm}{
\caption{Two different regions: Cronin enhancement due to interaction in the final state: a strong indication of the 
quark-gluon plasma production; and a quantum evolution region predicted bu the CGC  approach \cite{KLM} .}}
\end{minipage}
\label{npda3}
\end{figure}

I think that the situation with a suppression is clearly illustrated by \fig{npda3}. In this figure one can see two clearly 
separated regions with quite different physics. This separation suggests the way out: for the central and nuclear 
fragmentation region the anomalous dimension of the gluon density is not essential while in the forward fragmentation we see 
the effect of quantum evolution.

 \section{Three assumptions: a new edition}
However, the fact that we do not have a suppression in  the range of rapidities in central and nuclear fragmentation region, 
makes incorrect our beautiful explanation of the $N_{part}$ scaling for ion-ion collisions. We have to assume that this 
scaling stems not from initial condition for our evolution but rather from strong interaction in the final state  which 
suppress the production of the hadron from inside the nucleus. Only production from the nucleus surfaces can go out of the 
interacting system and can be observed. In this case we also have a $N_{part}$ scaling. Why we see this scaling up to $p_t 
\approx 5\,GeV$ ? We have not reached a clear understanding why. 

\begin{figure}[h]
\begin{center}
\includegraphics[width=0.90\textwidth]{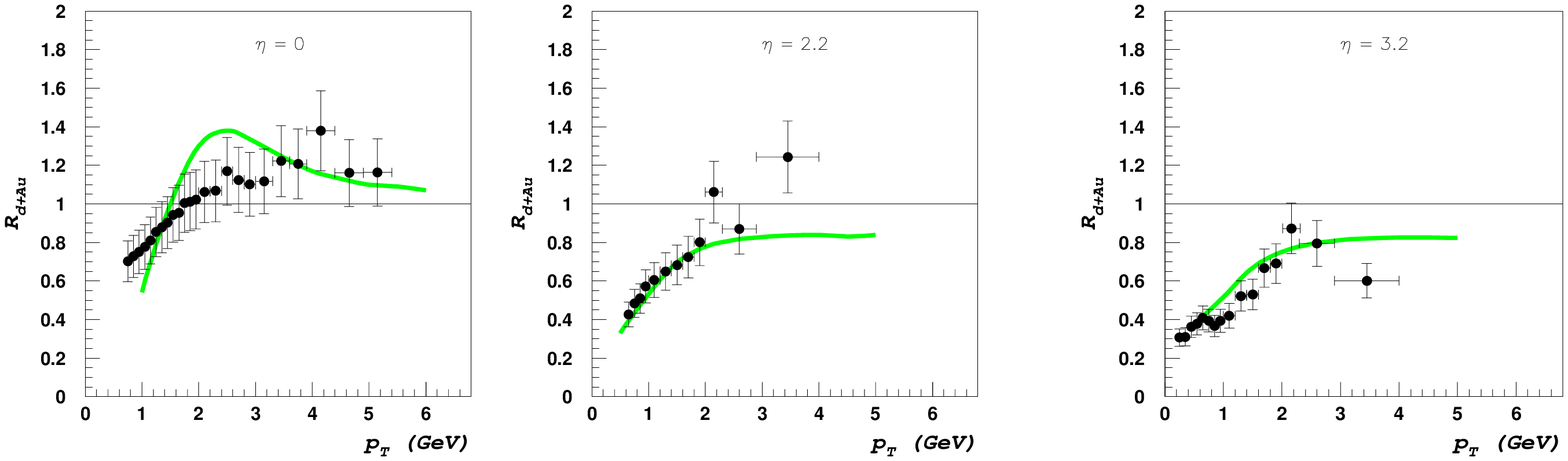}
\end{center}
\caption{The CGC approach, based on a new edition of our three assumptions, for deuteron-nucleus collisions \cite{KKT}. }
\label{dacgc}
\end{figure}

I hope that you understand that we have to change our three main assumptions under the press of the experimental data. Their 
new edition looks as follows:
\begin{enumerate}
\item \quad At $x \,>\,x_0\,\approx\,\,10^{-2}$ we have the McLerran-Venugopalan
model for the inclusive production of gluons with the saturation scale
$Q^2_s(x_0) \,\,\approx\,\,1\,GeV^2$;
\item \quad  $x\,\,<\,\,x_0\,\approx\,\,10^{-2}$ is low $x$, i.e. $\alpha_S \ln(1/x)
\,\,\approx\,\,1$ while $\alpha_S\,\,<\,\,1$;
\item \quad The  CGC theory determines the initial condition for the evolution of the partonic system in ion-ion collisions.  
The CGC 
 is the source of the thermalization which occurs at rather short
distance of the order of $1/Q_s$. Our formula is
{\it
CGC ( saturation)  is the initial condition  for hydrodynamical evolution in A
+ A
collisions.};
\end{enumerate}
\fig{dacgc} shows that the CGC approach is able to describe the experimental data using these three assumptions (see Ref. 
\cite{KKT}).

\begin{figure}[h]
\begin{center}
\includegraphics[width=0.90\textwidth]{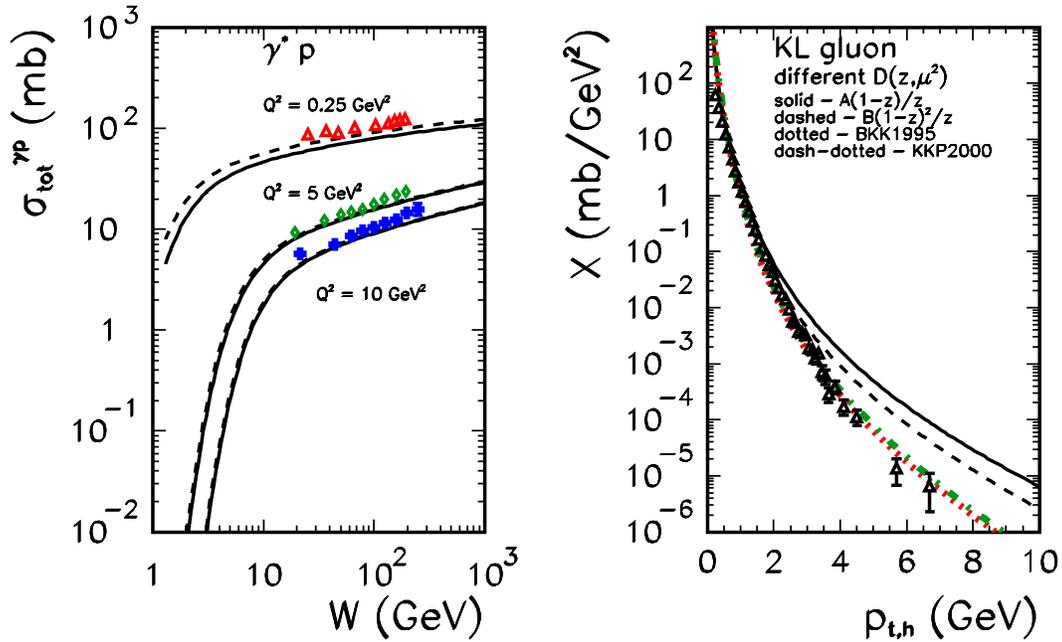}
\end{center}
\caption{ $p_t$ and $Q^2$ distribution in proton - proton collisions and DIS with the proton target. Pictures are taken from 
Ref. \cite{scz}. }
\label{ptpp1}
\end{figure}

\begin{boldmath}
 \section{$p_t$ distribution: proton-proton collisions.}
\end{boldmath}
The transverse momentum distribution is a very sensitive check of the form of our input for the parton densities.  In Ref. 
\cite{scz}  is shown that \eq{XGSAT} perfectly describe the $p_t$ and $^2$ distribution in proton-proton collision and DIS 
with the proton target (see \fig{ptpp1}). 
\begin{boldmath}
 \section{$p_t$ distribution: deuteron-nucleus  collisions.}
\end{boldmath}   

\begin{figure}[h]
\begin{center} 
\includegraphics[width=0.80\textwidth]{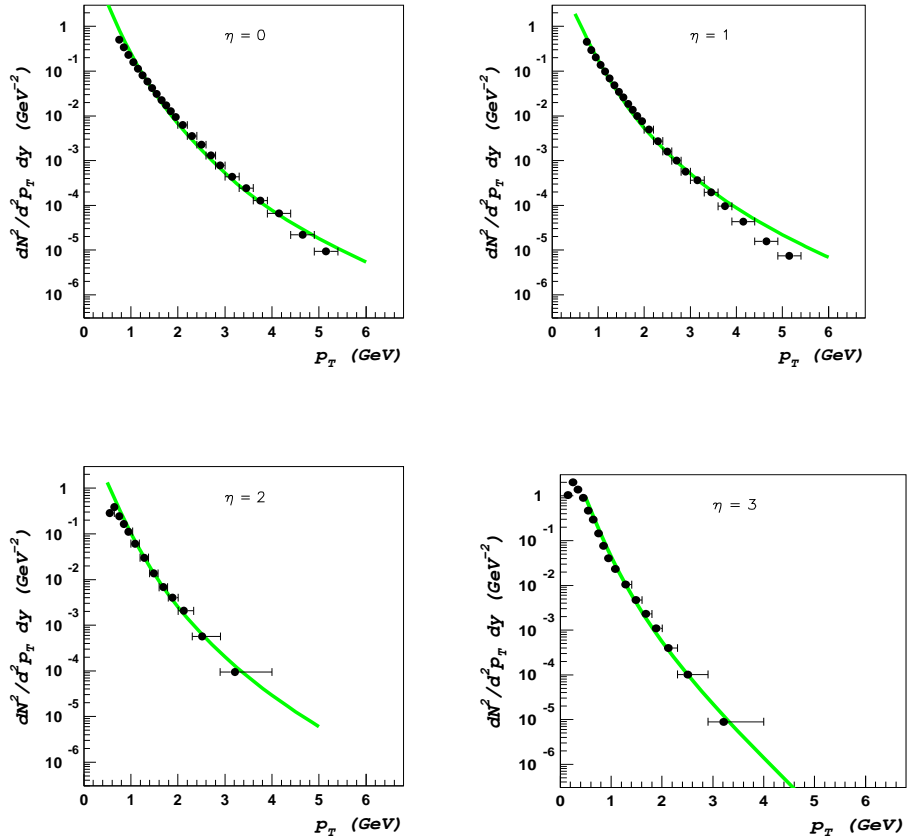}
\end{center}
\caption{ $p_t$  distribution in the deuteron - gold collisions in the CGC approach (see Ref. \cite{KKT}). }
\label{ptda}    
\end{figure}

In Ref. \cite{KKT} it is illustrated that the CGC approach is able to describe the transverse distributions of the produced 
hadrons in the deuteron-nucleus collisions n(see \fig{ptda}. This is a powerful check of our two first assumptions as well 
as the 
simplified form of the parton densities in \eq{XGSAT}.

\begin{figure}[h]
\begin{minipage}{18pc}
\includegraphics[width=18pc]{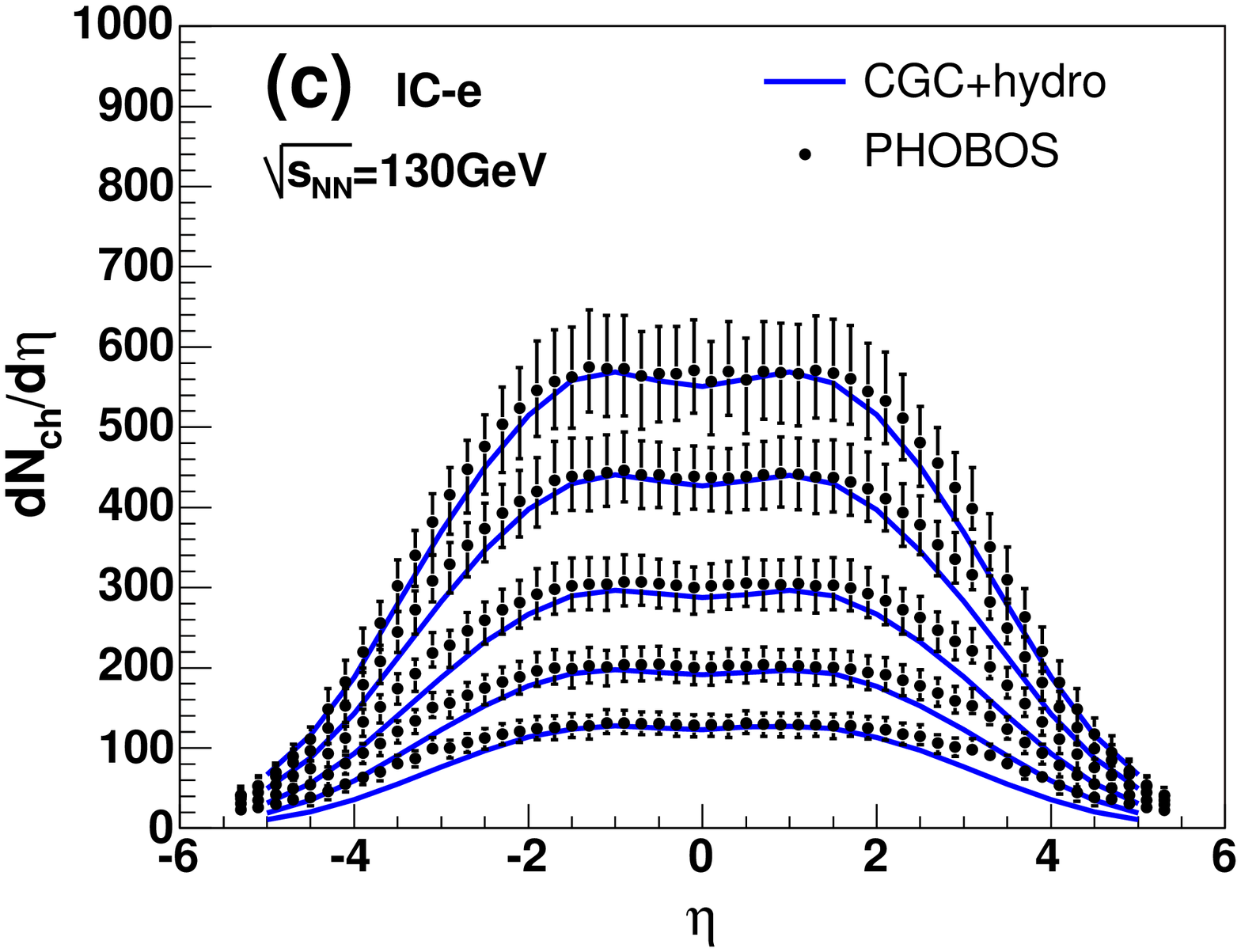}
\caption{\label{ptaa1}  Multiplicity distribution in CGC + hydro model of Ref.\cite{Hirano}}
\caption{\label{ptaa2} $p_t$  distribution in CGC + hydro model of Ref. \cite{Hirano}.}
\end{minipage}\hspace{2pc}%
\begin{minipage}{18pc}
\includegraphics[width=18pc]{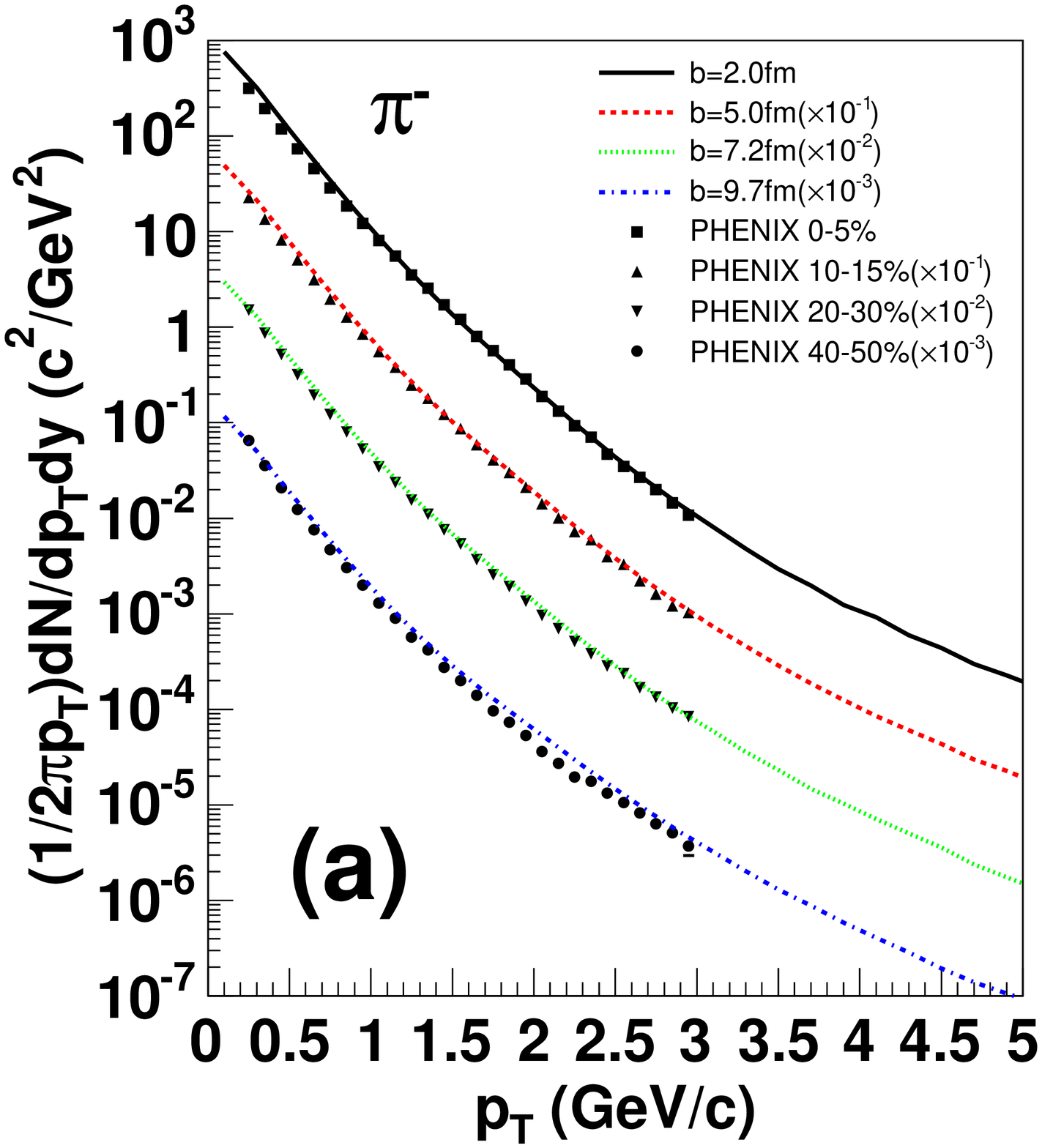}
\end{minipage}
\end{figure}

 \begin{boldmath}
 \section{$p_t$ distribution: ion-ion  collisions.}
\end{boldmath}
   In Ref. \cite{Eskola,Hirano} the good agreement with the experimental data is  achieved using the CGC theory as the 
initial condition for evolution which is treated in hydrodynamics (see \fig{ptaa1} and \fig{ptaa2}). This gives a strong 
argument in the 
support of out third assumption.

\section{Azimuthal correlations}
\begin{figure}[h]
\begin{minipage}{18pc}
\includegraphics[width=18pc]{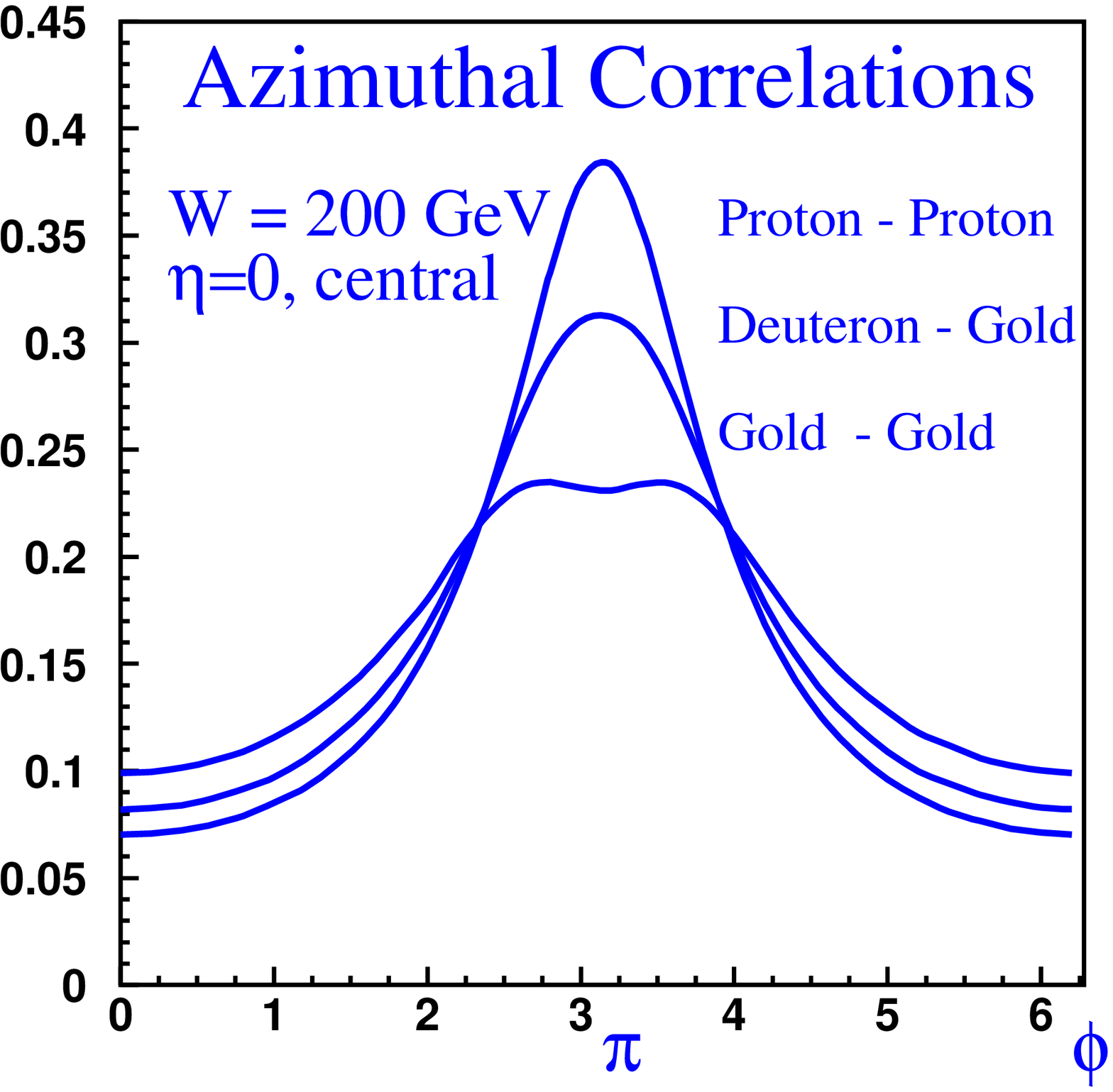}
\caption{\label{az1}  Our estimates \cite{KLMAZ} of the azimuthal correlations in one partons shower
for proton-proton, deuteron-gold and gold -gold collisions.}
\end{minipage}\hspace{2pc}%
\begin{minipage}{18pc}
\includegraphics[width=16pc]{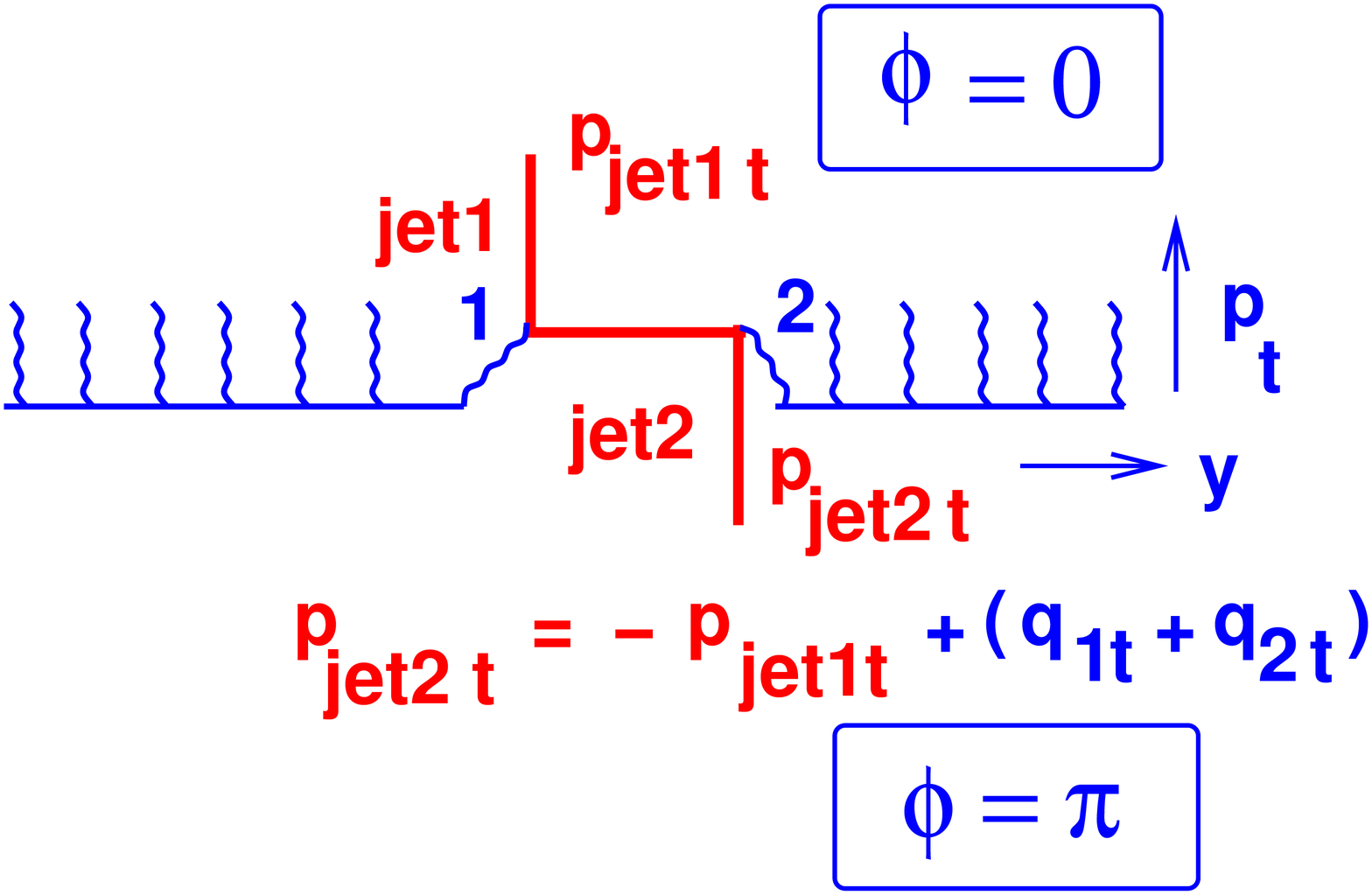}
\caption{\label{az2} Picture of the strong correlation in one parton shower.}
\end{minipage}
\end{figure}

\begin{figure}[h]
\begin{minipage}{18pc}
\includegraphics[width=16pc]{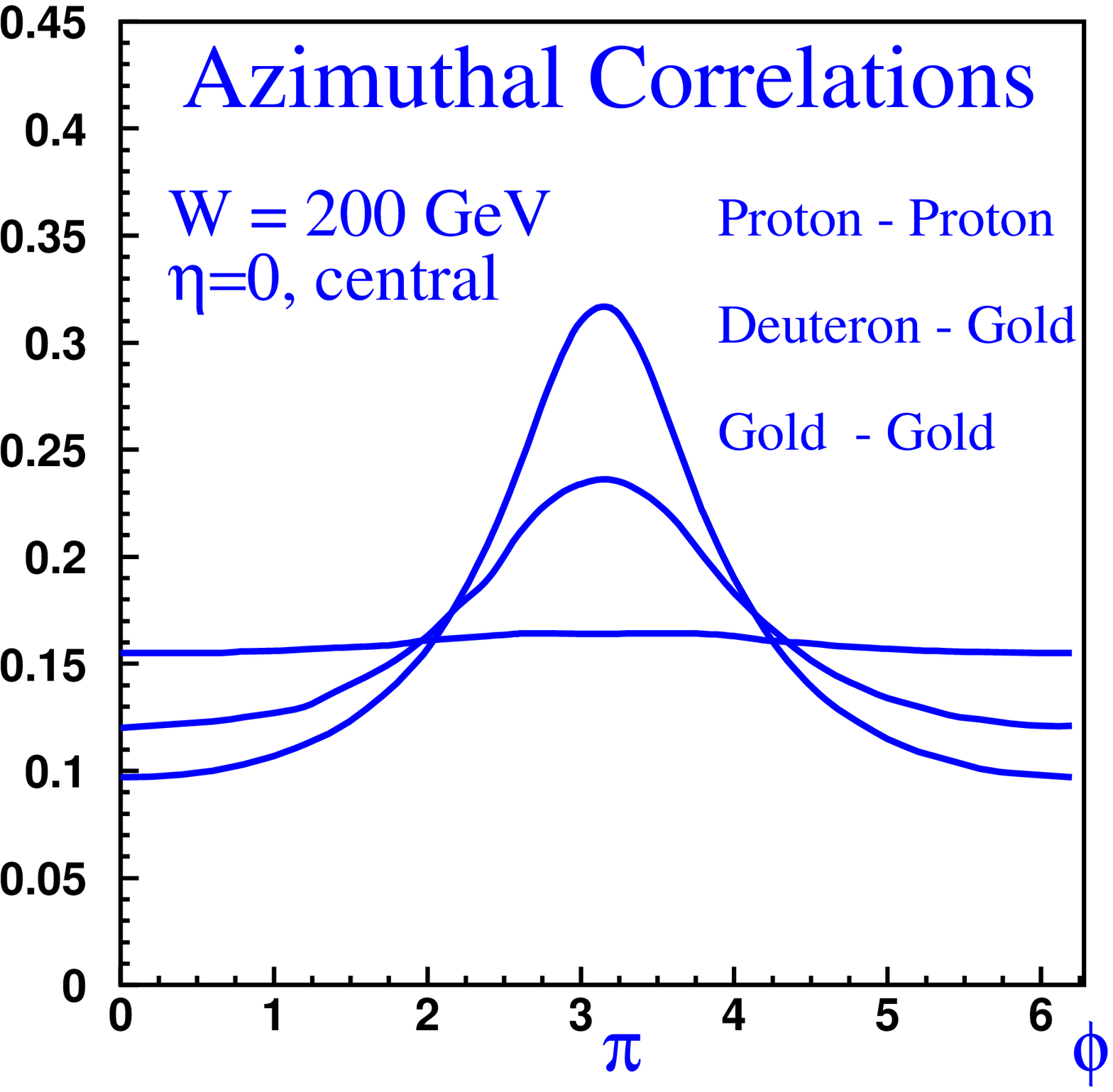}
\caption{\label{az3}  Our estimates \cite{KLMAZ} of the azimuthal correlations in one partons shower and in two parton 
showers (or in other words, due to independent particle production)
for proton-proton, deuteron-gold and gold -gold collisions.}
\end{minipage}\hspace{2pc}%
\begin{minipage}{18pc}
\includegraphics[width=18pc]{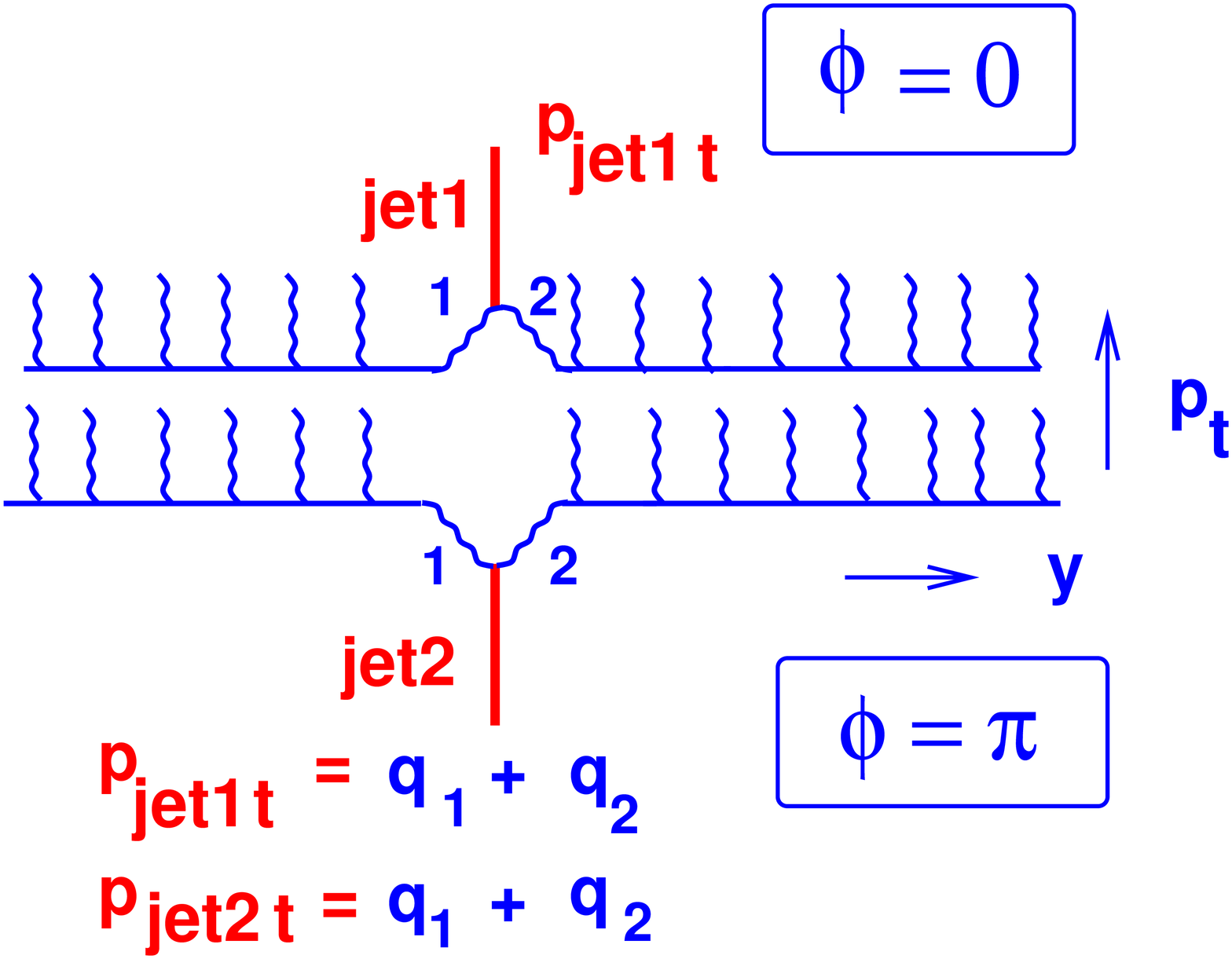}
\caption{\label{az4} Decorrelations in two parton showers.}
\end{minipage}
\end{figure}

The last issue that I want to touch in this talk is the azimuthal correlations since they give a beautiful example of our 
expectation in the CGC approach. In the new phase-CGC, we expect  decorrelations mostly because the source of the 
production of two particles with definite azimuthal angle between them is just independent production of two hadrons (see 
\fig{az4}). Such a 
production is proportional to the square of parton density and it gives a small contribution in pQCD phase of QCD. In this 
phase the typical process is back-to-back correlation due to the hard rescattering of two partons that belong to the same 
parton shower (see \fig{az2}). 

\begin{figure}[h]
\begin{center}
\includegraphics[width=0.70\textwidth]{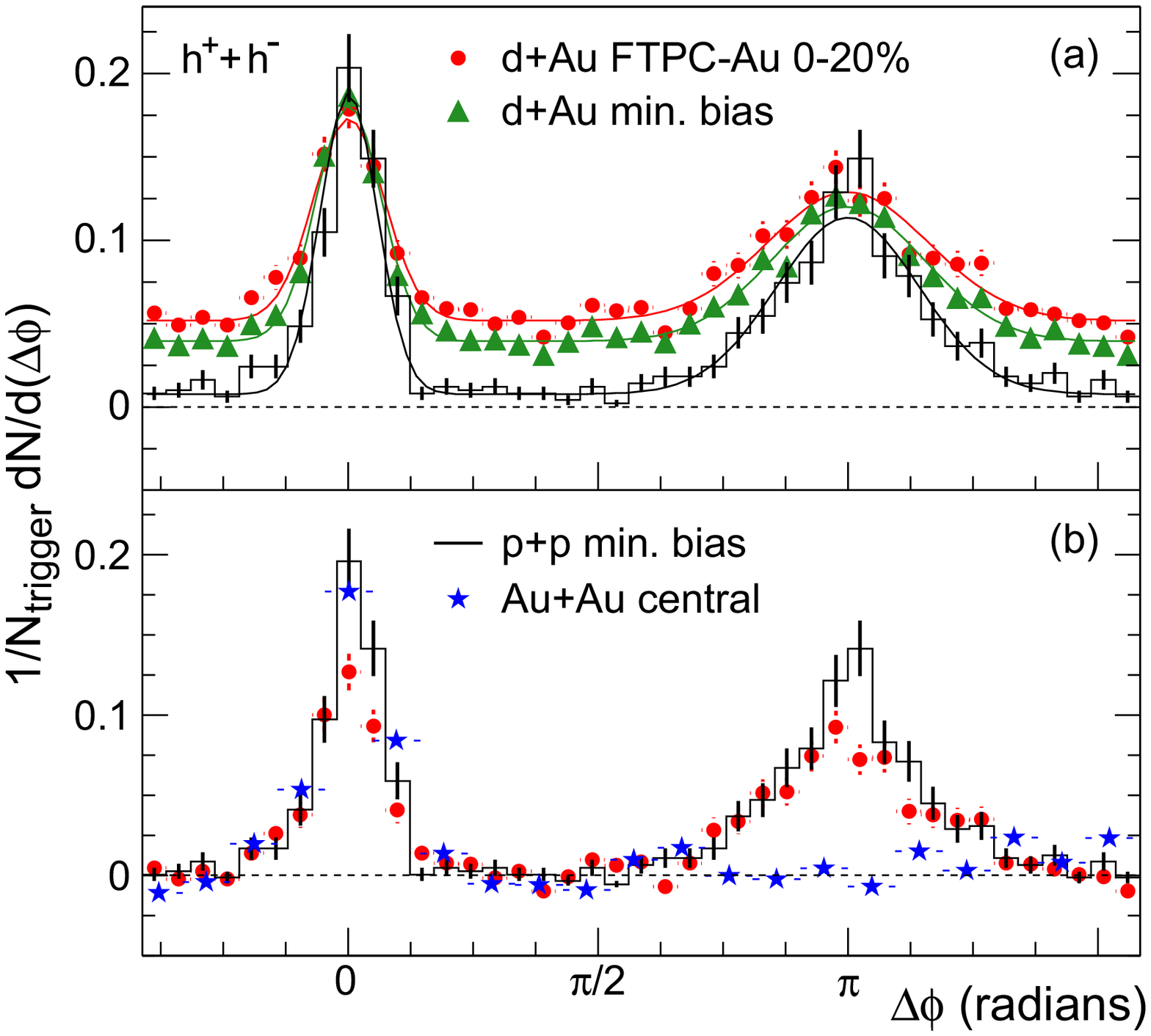}
\end{center}
\caption{Star data for azimuthal correlations. }
\label{az5} 
\end{figure}

\fig{az1} and \fig{az3} show our estimates \cite{KLMAZ} of the correlation in one parton shower (\fig{az1}) and  the 
resulting correlations (\fig{az3}) including the production from two parton showers.
Our estimates certainly reproduce the experimental data of Star collaboration ( see \fig{az5}).

\section{Resume}
I hope that I convinced you that the distance between experimental data and our microscopic theory: QCD, is not so long as 
usually people think. The RHIC experimental data  support the idea that at high parton density we are dealing with a new 
phase - Colour Glass Condensate. The parton density saturation, the most essential property of the CGC 
phase , manifests itself in the RHIC data. The most essential characteristics of the data that can be easily and elegantly 
described in the CGC approach are (i) energy, rapidity and centrality dependence of the particle multiplicities; (ii) the 
suppression of the rapidity distribution in the deuteron-nucleus collisions  in comparison with the proton-proton collisions 
in forward 
region of rapidity; (iii) suppression at $\eta=0$ for rather small transverse momenta in the same data and (iv) observed 
azimuthal decorrelation in gold-gold collisions.

I hope that future experiments at RHIC and LHC will provide more arguments in favour of the CGC and our understanding of main 
features of QCD will become deeper and more transparent.

\end{document}